\documentclass[prb,twocolumn,showpacs]{revtex4}
\usepackage{graphicx,amsmath,amssymb}
\begin{document}

\renewcommand{\Re}{\mathop{\mathrm{Re}}}
\renewcommand{\Im}{\mathop{\mathrm{Im}}}
\renewcommand{\b}[1]{\mathbf{#1}}
\renewcommand{\c}[1]{\mathcal{#1}}
\renewcommand{\u}{\uparrow}
\renewcommand{\d}{\downarrow}
\renewcommand{\mod}{\mathop{\mathrm{mod}}}
\newcommand{\bsigma}{\boldsymbol{\sigma}}
\newcommand{\blambda}{\boldsymbol{\lambda}}
\newcommand{\tr}{\mathop{\mathrm{tr}}}
\newcommand{\sgn}{\mathop{\mathrm{sgn}}}
\newcommand{\sech}{\mathop{\mathrm{sech}}}
\newcommand{\diag}{\mathop{\mathrm{diag}}}
\newcommand{\half}{{\textstyle\frac{1}{2}}}
\newcommand{\quarter}{{\textstyle\frac{1}{4}}}
\newcommand{\sh}{{\textstyle{\frac{1}{2}}}}
\newcommand{\ish}{{\textstyle{\frac{i}{2}}}}
\newcommand{\thf}{{\textstyle{\frac{3}{2}}}}
\newcommand{\D}{\mathcal{D}}

\title{
Field theory of the quantum Hall nematic transition}

\author{J. Maciejko$^1$, B. Hsu$^2$, S. A. Kivelson$^3$, YeJe Park$^2$, and S. L. Sondhi$^2$}

\affiliation{$^1$Princeton Center for Theoretical Science, Princeton University, Princeton, New Jersey 08544, USA\\ $^2$Department of Physics, Princeton University, Princeton, New Jersey 08544, USA\\
$^3$Department of Physics, Stanford University, Stanford, California 94305, USA
}

\date\today

\begin{abstract}

The topological physics of quantum Hall states is efficiently encoded in 
purely topological quantum field theories of the Chern-Simons type. The
reliable inclusion of low-energy dynamical properties in a continuum 
description however typically requires proximity to a quantum critical point.
We construct a field theory that describes the quantum transition from an isotropic to a nematic Laughlin liquid.  The soft mode associated with this transition approached from the isotropic side is identified as the familiar intra-Landau level Girvin-MacDonald-Platzman mode. We obtain $z=2$ dynamic
scaling at the critical point and a description of Goldstone and defect
physics on the nematic side. Despite the very different physical motivation,
our field theory is essentially identical to a recent ``geometric'' field 
theory for a Laughlin liquid proposed by Haldane.

%

\end{abstract}


\maketitle

\section{Introduction}

The problem of constructing a 
 field theoretic description
of fractional quantum Hall (FQH) states has a venerable history. The challenge was posed by
Girvin,\cite{QHE} important progress was made by Girvin and MacDonald\cite{girvin1987} and then the challenge was substantially met via the construction of the bosonic\cite{zhang1989} and
fermionic\cite{lopez1991} Chern-Simons (CS) theories of FQH states or equivalently
their composite boson and fermion descriptions. These constructions, however,
have remained somewhat unsatisfying in their ability to reproduce the
high magnetic field limit where one expects to see the dynamics manifestly
projected onto the lowest Landau level. Recently, Haldane has proposed
an alternative field theory involving a fluctuating unimodular
``metric'' field $g_{ab}(\b{r},t)$ which is intended to fix these problems.\cite{haldane2011,HaldaneFieldTheory}

In this paper we also revisit the challenge of giving a field theoretic
description of the QH effect but from {\it prima facie} a very different but much more conventional viewpoint wherein a field theory arises in the proximity
of a continuous transition between two phases. To this end we consider a
transition between an isotropic Laughlin liquid at filling factor $\nu=1/q$
and a hypothesized QH nematic at the same $\nu$ which exhibits the QH effect 
but spontaneously breaks rotational symmetry.\cite{unquantizedFQHN} [We note that such a phase has been recently proposed in Ref.~\onlinecite{mulliganFQHN} motivated by recent experiments\cite{xia2011} and a simpler version, the QH
Ising nematic proposed in Ref.~\onlinecite{abanin2010}, has also very likely already been seen in experiments.\cite{shkolnikov2005,gokmen2010}]
%
We construct a field theory for
the isotropic to nematic transition which involves the QH
order coupled to the traceless, symmetric nematic matrix order
parameter $\hat{Q}_{ab}(\b{r},t)$.\cite{DeGennes} This construction closely
parallels the field theoretic treatment of QH ferromagnetism\cite{sondhi1993}
and
has a mathematically elegant relation to it analogous to that of Minkowski
and Euclidean spaces.

The resulting field theory on the isotropic side of the transition is
%
 %
 essentially
identical to Haldane's upon the natural identification
\begin{align}
g=\exp\hat{Q},\nonumber
\end{align}
via a matrix exponentiation.
%
%
Building on the early suggestion of Zhang and Lee\cite{lee1991} and the results of the recent
detailed study by Yang \emph{et al.}\cite{yang2012} that the famous Girvin-MacDonald-Platzman (GMP) mode\cite{GMP} of the Laughlin liquid is a fluctuating quadrupole at long wavelengths, we identify it as the precursor 
of the mode that eventually goes soft at the nematic transition (assuming 
it is continuous).  However, the Laughlin quasiparticles {\it do
not} become gapless at the transition and they are vortices/fluxons of the CS
field rather than solitons of the order parameter. Likewise the long-wavelength Kohn
mode decouples from the low-energy physics of the transition.

Our field theory describes
a $z=2$ isotropic-to-nematic transition in the universality class of the 2D dilute Bose gas and enables us to compute the
collective mode spectrum and electromagnetic response of the nematic phase. We find a linearly dispersing neutral Goldstone mode and a quantized Hall conductivity. Interestingly, a critical point with $z=2$ scaling was also found in Ref.~\onlinecite{mulliganFQHN}, but in a very different setting where the spontaneous breaking of rotational symmetry is not described by the conventional nematic order parameter $\hat{Q}_{ab}$ but rather occurs in the CS gauge field sector.

Before proceeding to the technical content of this paper we would like to briefly
recapitulate the conceptual framework underlying continuum limits in broken symmetry
systems and discuss its extension to systems with topological order, such as the FQH effect. In systems where ordering proceeds via symmetry breaking we can obtain
continuum limits 
via two
distinct routes.
The first involves phases that exhibit Goldstone modes and hence are {\it generically} gapless
and exhibit 
algebraic correlations. Here the field theories
arise in a scaling limit where all distances/times are large on the microscopic scale anywhere
in the phase.
The second route is to work near a critical (or multicritical) point, typically
between an ordered and symmetric phase in a scaling limit where all distances/times are fixed
multiples of the correlation length/time which are tuned to be large on microscopic scales.
The first route naturally yields sigma models which encapsulate Goldstone physics as well as the
physics of topological defects while the second naturally yields the Landau-Ginzburg-Wilson (LGW) field theories for
understanding critical behavior. The point worth noting is that in order to describe
anything other than the topological properties of gapped
phases---either symmetric or with broken discrete symmetries---we are forced to take the
second approach.

Turning now to topologically ordered phases, let us loosely define them as phases which exhibit
emergent low-energy gauge fields. Here again we can consider cases where there are generic
gapless excitations, e.g., photons in (3+1)-dimensional quantum Coulomb phases or in (2+1)-dimensional algebraic spin liquids which exhibit low-energy gauge and matter degrees of freedom.
But Laughlin liquids are topologically ordered in the original, narrower sense. They are gapped
but their scaling limits, constructed by taking all distances/times large on the microscopic
scale, are however nontrivial and described by topological quantum field theories which lack
low energy and local bulk degrees of freedom.\cite{WenBook} Specifically
for the Laughlin liquids the CS theory encodes their electromagnetic response, ground
state degeneracies on closed manifolds, quasiparticle braiding and the symplectic structure
for their edge states (which are now again gapless degrees of freedom and thus 
amenable
to 
a field theoretic treatment).

If we wish to move beyond the purely topological description of the QH states and other
phases that exhibit gapped topological order, it is now necessary to 
find a regime in which fluctuations
occur on
long length 
and low energy scales, exactly as in the case of gapped phases in the LGW paradigm.
What is new here is that topologically ordered phases are stable to {\it all} local perturbations
so it is possible to imagine multiple perturbations that might accomplish this corresponding
to continuous transitions to different neighboring phases. In this sense there are presumably
multiple field theories that yield descriptions of topological phases beyond the purely topological
scaling limit. For example, the Haldane phase\cite{HaldaneGap} of the antiferromagnetic $S=1$ spin chain, which is a symmetry-protected topological phase,\cite{gu2009,pollmann2012} is described in the continuum by the disordered phase of the $O(3)$ nonlinear sigma model\cite{HaldaneGap} (NL$\sigma$M) as well as by a massive deformation of the $SU(2)_2$ Wess-Zumino-Witten conformal field theory.\cite{affleck1986} The study we undertake here allows us to obtain
%
a field
theory for the Laughlin liquids from a particular phase transition, that en route to a QH nematic. As the transition involves the onset of a broken symmetry, the exercise involves
a piece of canonical LGW analysis with the additional feature that the order parameter introduced
into the problem has to be appropriately coupled to the existing topological gauge field.
As applied to the isotropic phase, this approach defines a limit in which the GMP mode has a field theoretic description.
One could also imagine taking a similar approach to elucidating the origin of  the magnetoroton minimum, first found employing the same single-mode approximation which led to the discovery of the GMP mode.\cite{GMP} Intuitively, this has already been interpreted as a signature of an incipient transition to a Wigner crystalline phase.  However, as the Wigner crystalline transition is generically expected to be first order, it is a bit subtle to define a field theory limit in which this identification could be made precise;  possibly if one considered the problem in a system with explicit rotational symmetry breaking (where the transition to a translation symmetry breaking state could be continuous), an analogous approach to the one taken here could yield a more rigorous understanding of this mode. Finally we note interesting early work on many-electron wave functions\cite{musaelian1996} as well as a field theory\cite{balentshexatic} for spatially ordered QH phases, although with significant differences to the approach taken here.

This paper is structured as follows. In Sec.~\ref{sec:QHOP}, we review the basic ingredients of a field-theoretic description of the FQHE, restricting ourselves to the simplest case of the Abelian Laughlin FQH liquids at filling fraction $\nu=1/q$, with $q$ an odd integer. In Sec.~\ref{sec:OPinternal}, we introduce an order parameter for the spontaneous breaking of an internal symmetry and describe our prescription for coupling this order parameter to the topological degrees of freedom of the FQHE. We then illustrate this general procedure in Sec.~\ref{sec:QHFM} with the example of QH ferromagnetism. In Sec.~\ref{sec:QHN}, we follow a similar approach to construct a theory of the FQH isotropic-to-nematic transition and comment in some detail on the
connection to Haldane's field theory. 
We adapt the formalism of Sec.~\ref{sec:OPinternal} to deal with the spontaneous breaking of rotation symmetry, a spatial symmetry. The resulting field theory turns out to be a $SO(2,1)$ or ``Minkowski'' analog of the $SO(3)$ QH ferromagnetism problem. We make some concluding remarks and summarize the
contents of the paper in Sec.~\ref{sec:conclusion}.

\section{Field theory description of the quantum Hall liquid}
\label{sec:QHOP}

The topological scaling limit of the $\nu=1/q$ Laughlin state is captured by the topological field theory\cite{zhang1989}
\begin{align}\label{Ltop}
\c{L}&=\frac{1}{4\pi q}\epsilon^{\mu\nu\lambda}
\alpha_\mu\partial_\nu\alpha_\lambda
-J^\mu(\partial_\mu\theta+\alpha_\mu+A_\mu),
\end{align}
where $A_\mu$ is the external electromagnetic field, $\alpha_\mu$ is an emergent dynamical gauge field, and $J^\mu=(\rho,\b{J})=-\partial\c{L}/\partial A_\mu$ is a matter field corresponding to the electric charge density $\rho$ and current $\b{J}$. The compact $U(1)$ variable $\theta$ can be decomposed into smooth and vortex parts. The vortex part quantizes the fluxes of $\boldsymbol{\alpha}$ via the equation of motion for $\b{J}$ which implies $\nabla\times(\boldsymbol{\alpha}+\b{A})
=\nabla\times\nabla\theta$. The smooth part can be integrated out to impose current conservation $\partial_\mu J^\mu=0$. The theory (\ref{Ltop}) has a well-known dual,
\cite{fisher1989}
  obtained by writing the conserved current as $J^\mu=\frac{1}{2\pi}\epsilon^{\mu\nu\lambda}\partial_\nu a_\lambda$ with $a_\mu$ a dual gauge field and integrating out $\alpha_\mu$,
\begin{align}
\c{L}_\textrm{dual}=\frac{q}{4\pi}\epsilon^{\mu\nu\lambda}
a_\mu\partial_\nu a_\lambda-J^\mu_\textrm{vor}a_\mu
-\frac{1}{2\pi}\epsilon^{\mu\nu\lambda}
A_\mu\partial_\nu a_\lambda,\nonumber
\end{align}
where $J^\mu_\textrm{vor}=\frac{1}{2\pi}
\epsilon^{\mu\nu\lambda}\partial_\nu\partial_\lambda
\theta_\textrm{vor}$ is the Laughlin quasiparticle current. The dual gauge field $a_\mu$ can then be integrated out to yield the fractional quantized Hall conductivity, in the absence of Laughlin quasiparticles, or fractional statistics for the quasiparticles, in the absence of external electromagnetic fields. In this purely topological description, there are no energies associated with the electromagnetic or quasiparticle currents.

To go beyond the topological scaling limit, a Hamiltonian should be added. From a hydrodynamic point of view, there should be an energy cost associated with departures of the density $\rho$ and current $\b{J}$ from their ground state values. We consider the simple Hamiltonian
\begin{align}
\c{H}_\textrm{int}(J^\mu)=\frac{u}{2}(\rho-\bar{\rho})^2
-\frac{1}{2\kappa\rho}\b{J}^2,\nonumber
\end{align}
where $u$, $\kappa$, and $\bar{\rho}$ are positive constants. The first term describes repulsive density interactions and
favors $\rho=\bar{\rho}$.
 For simplicity we consider short-range interactions, but the Coulomb interaction could have been chosen instead. The second term sets the current equal to $\b{J}=\kappa\rho
(\nabla\theta-\boldsymbol{\alpha}-\b{A})$, and
favors $\b{J}=0$.
For a translationally invariant ground state $\nabla\theta=0$, this implies $\boldsymbol{\alpha}+\b{A}=0$ which, in conjunction with the equation of motion for $\alpha_0$, gives the correct Laughlin filling $\bar{\rho}=\frac{\nu}{2\pi\ell_B^2}$ with $\ell_B=\sqrt{1/B}$ the magnetic length.

The above theory\cite{zhang1989} reproduces the gapped cyclotron or Kohn mode with dispersion $\varepsilon(\b{q})=\kappa B+\c{O}(q^2)$ and gives a finite energy to the Laughlin quasiparticles.\cite{zhang1989,tafelmayer1993} It has the disadvantage that the quasiparticle energy depends on the parameter $\kappa$ which, according to Kohn's theorem,\cite{kohn1961} should depend only on the bare electron mass which is manifestly not a property of the lowest Landau level. Although our construction does not solve this problem, as we will see the Laughlin quasiparticles remain gapped through the isotropic-to-nematic transition and into the QH nematic phase. As a result, they affect the low-energy and long-wavelength properties of neither. On the other hand, the GMP mode is identified with fluctuations of the nematic order in the isotropic phase, and its energy scale is set by a parameter in our effective field theory which is unrelated to the bare electron mass.

\section{Spontaneous breaking of an internal symmetry}
\label{sec:OPinternal}

We now introduce an order parameter for a phase transition out of the FQH state. In this section we focus on the spontaneous breaking of an internal symmetry, for example, spin rotation symmetry which will be discussed in Sec.~\ref{sec:QHFM}. In Sec.~\ref{sec:QHN} we discuss the spontaneous breaking of a spatial symmetry such as rotation symmetry. In general, the order parameter for an internal symmetry is a multi-component scalar field $\boldsymbol{\Phi}=(\Phi^1,\Phi^2,\ldots)$ which lives in some target space $M$. From a field theory standpoint, we need to write down a Lagrangian for $\boldsymbol{\Phi}$ itself, and couple $\boldsymbol{\Phi}$ to the field theory just discussed for an isotropic FQH liquid. The static term in the Lagrangian for $\boldsymbol{\Phi}$ is easy to write down and corresponds to the free energy density $\c{H}_\textrm{OP}(\boldsymbol{\Phi},\partial_i\boldsymbol{\Phi})$ for a classical phase transition. The time-dependent term defines the quantum dynamics of $\boldsymbol{\Phi}$ and should be consistent with the broken time-reversal symmetry of the FQH liquid. The degrees of freedom of the isotropic FQH liquid are the electromagnetic current $J^\mu$ and CS gauge fields $\alpha_\mu$ or $a_\mu$, and $\boldsymbol{\Phi}$ should couple to these in some way.

We now describe an approach which allows us to solve the problem of the quantum dynamics of $\boldsymbol{\Phi}$ and its coupling to the FQH liquid in a mathematically elegant way. We assume that the target space $M$ can be written in the form $M=G/H$ where $G$ is a Lie group and $H$ is a closed Lie subgroup of $G$. When $G$ is compact, the field theory we construct describes the transition from the point of view of the ordered phase, and $M$ should be interpreted as the Goldstone manifold for a transition where $G$ is spontaneously broken to $H$. A typical example is the $O(N)$ NL$\sigma$M for magnetic transitions.\cite{Sachdev} When $G$ is noncompact, the field theory is a Landau-Ginzburg theory for the transition. The advantage of writing $M$ in this form is that it allows for the construction of a new gauge field $\alpha_\mu^\textrm{OP}$ from the degrees of freedom of the order parameter itself,
\begin{align}
\alpha_\mu^\textrm{OP}=\alpha_\mu^\textrm{OP}(\boldsymbol{\Phi},\partial_\nu\boldsymbol{\Phi}),\nonumber
\end{align}
which depends only on $\boldsymbol{\Phi}$ and its derivatives. The gauge group for $\alpha_\mu^\textrm{OP}$ is $H$. This construction is based on the mathematical fact that $G$ can be viewed as a principal $H$-bundle over $M=G/H$,\cite{nakahara} and this bundle admits a connection 1-form $\c{A}$ valued in the Lie algebra of $H$. Promoting the order parameter $\boldsymbol{\Phi}\in M$ to a dynamical field $\boldsymbol{\Phi}(x^\mu)$ in $(2+1)$D spacetime, we can pull back the connection 1-form $\c{A}=\c{A}_I\textrm{d}\Phi^I$ on $M$ to obtain a gauge field $\alpha_\mu^\textrm{OP}(x^\nu)$ in $(2+1)$D spacetime,
\begin{align}\label{OPgaugefield}
\alpha_\mu^\textrm{OP}(x^\nu)=\frac{\partial\Phi^I}{\partial x^\mu}\c{A}_I(\Phi^J(x^\nu)).
\end{align}
For applications to Abelian FQH states which are described by $U(1)$ gauge theories, we are thus led to study target spaces of the form $M=G/U(1)$.

To describe a transition in the FQH liquid, we couple the order parameter gauge field $\alpha_\mu^\textrm{OP}$ to the electromagnetic current $J^\mu$ in the FQH topological Lagrangian (\ref{Ltop}) augmented by $\c{H}_\textrm{int}(J^\mu)$ and the static free energy density $\c{H}_\textrm{OP}(\boldsymbol{\Phi},\partial_i\boldsymbol{\Phi})$. We will henceforth refer to this approach as the direct approach. Our effective field theory is obtained by integrating out $\b{J}$,
\begin{align}\label{LOP}
\c{L}=&-\frac{u}{2}(\rho-\bar{\rho})^2
-\rho(\partial_0\theta+\alpha_0^\textrm{OP}+\alpha_0+A_0)\nonumber\\
&-\frac{\kappa\rho}{2}(\partial_i\theta+\alpha_i^\textrm{OP}
+\alpha_i+A_i)^2
+\frac{1}{4\pi q}\epsilon^{\mu\nu\lambda}
\alpha_\mu\partial_\nu\alpha_\lambda\nonumber\\
&-\c{H}_\textrm{OP}(\boldsymbol{\Phi},\partial_i\boldsymbol{\Phi}).
\end{align}
In this approach, the FQH state is a mean-field state corresponding to the superfluid phase of Eq.~(\ref{LOP}) with $\bar{\rho}>0$.\cite{zhang1989} In the temporal gauge $A_0=0$, there is a uniform, static solution to the classical equations of motion derived from Eq.~(\ref{LOP}), which is given by
\begin{align}\label{FQHOPsolution}
\alpha_0=0,\hspace{5mm}
\boldsymbol{\alpha}=-\b{A},\hspace{5mm}
\rho=\bar{\rho},\hspace{5mm}
\boldsymbol{\Phi}=\textrm{const.}
\end{align}
Because $\boldsymbol{\Phi}$ is static and uniform, Eq.~(\ref{OPgaugefield}) implies that $\alpha_\mu^\textrm{OP}=0$. To obtain a low-energy effective field theory of the FQHE, we consider small long-wavelength fluctuations about the mean-field solution (\ref{FQHOPsolution}),
\begin{align}
\delta\alpha_\mu=\alpha_\mu+A_\mu,\hspace{5mm}
\delta\rho=\rho-\bar{\rho},\hspace{5mm}
\boldsymbol{\Phi}\neq\textrm{const.},\nonumber
\end{align}
where small long-wavelength fluctuations of $\boldsymbol{\Phi}$ imply that $\alpha_\mu^\textrm{OP}$ is small. Keeping only terms up to quadratic order in the fluctuations, we obtain
\begin{align}\label{LagrangianQHOPdirect}
\c{L}=&-\frac{u}{2}(\delta\rho)^2-\delta\rho
(\partial_0\theta+\alpha_0^\textrm{OP}+\delta\alpha_0)
\nonumber\\
&-\frac{\kappa\bar{\rho}}{2}(\partial_i\theta+\alpha_i^\textrm{OP}+\delta\alpha_i)^2
+\frac{1}{4\pi q}\epsilon^{\mu\nu\lambda}
\delta\alpha_\mu\partial_\nu\delta\alpha_\lambda
\nonumber\\
&+\c{L}_\textrm{OP}(\boldsymbol{\Phi},\partial_\mu\boldsymbol{\Phi}),
\end{align}
where the Lagrangian describing the dynamics of the order parameter is
\begin{align}
\c{L}_\textrm{OP}(\boldsymbol{\Phi},\partial_\mu\boldsymbol{\Phi})=-\bar{\rho}
\c{A}_I(\boldsymbol{\Phi})\partial_0\Phi^I
-\c{H}_\textrm{OP}(\boldsymbol{\Phi},\partial_i\boldsymbol{\Phi}),\nonumber
\end{align}
using Eq.~(\ref{OPgaugefield}). The quantum dynamics of the order parameter is determined by the first term which is odd under time reversal $t\rightarrow-t$. This reflects the macroscopic time-reversal symmetry breaking in the FQH state.

A dual description is obtained as previously,\cite{lee1990}
\begin{align}\label{FQHOPMCS}
\c{L}_\textrm{dual}=&\frac{q}{4\pi}\epsilon^{\mu\nu\lambda}a_\mu\partial_\nu a_\lambda
-(J^\mu_\textrm{vor}+J^\mu_\textrm{OP})a_\mu\nonumber\\
&-\frac{1}{2\pi}\epsilon^{\mu\nu\lambda}A_\mu\partial_\nu
a_\lambda
+\c{L}_\textrm{OP},
\end{align}
where $J^\mu_\textrm{OP}$ is an additional topological current constructed from the order parameter,
\begin{align}\label{OPtopcurrent}
J^\mu_\textrm{OP}&=\frac{1}{2\pi}\epsilon^{\mu\nu\lambda}\partial_\nu
\alpha_\lambda^\textrm{OP},
\end{align}
which describes topological excitations of the order parameter. The dual gauge field $a_\mu$ could then be integrated out from Eq.~(\ref{FQHOPMCS}) to yield a Hopf term for the total topological current $J^\mu_\textrm{vor}+J^\mu_\textrm{OP}$, implying $\pi/q$ fractional statistics for both Laughlin quasiparticles and topological excitations of the order parameter. In the absence of quasiparticles $J^\mu_\textrm{vor}+J^\mu_\textrm{OP}=0$, the order parameter decouples from the dual gauge sector. The latter is a Maxwell-Chern-Simons theory with nontrivial topological degeneracy on Riemann surfaces.\cite{wen1989,wen1990}

\subsection{Quantum Hall ferromagnetism}
\label{sec:QHFM}

The physics of QH ferromagnetism\cite{sondhi1993,barrett1995} can be understood as an example of the general approach just described. The order parameter in this case is an $SO(3)$ vector. Deep in the ordered phase, we can neglect the massive amplitude fluctuations and keep only the direction, which is described by a unit vector $\b{n}$ in 3D Euclidean space $\mathbb{R}^3$ with $\b{n}^2=1$. Therefore, the target space $M$ in which the order parameter $\b{n}$ lives is the 2-sphere $\mathbb{S}^2$, which can be written as\cite{nakahara}
\begin{align}\label{SO3SO2}
\mathbb{S}^2=SO(3)/SO(2),
\end{align}
where $SO(2)$ corresponds to rotations about $\b{n}$.

We first consider the direct approach, according to which we should construct a $SO(2)=U(1)$ gauge field from $\b{n}$. From a physical standpoint, there is a natural solution to this problem. The quantum dynamics of an $SO(3)$ rotor can be modeled by a charged particle moving on the surface of a sphere with a magnetic monopole located at the origin, where $\b{n}$ is the coordinate of that particle.\cite{Fradkin} The vector potential $\boldsymbol{\c{A}}$ associated with this magnetic monopole satisfies the equation
\begin{align}
\nabla_\b{n}\times\boldsymbol{\c{A}}(\b{n})=s\b{n},\nonumber
\end{align}
where $s$, the spin quantum number, is the magnetic charge of the monopole. A solution for $\boldsymbol{\c{A}}$ valid everywhere except at the south pole of $\mathbb{S}^2$ is the Wu-Yang potential,\cite{Sakurai}
\begin{align}
\boldsymbol{\c{A}}(\b{n})=s\left(\frac{1-\cos\vartheta}{\sin\vartheta}\right)\hat{\boldsymbol{\varphi}},\nonumber
\end{align}
where the embedding of $\b{n}$ in $\mathbb{R}^3$ is given by the usual spherical coordinates $\b{n}=(\sin\vartheta\cos\varphi,\sin\vartheta\sin\varphi,\cos\vartheta)$. The order parameter gauge field $\alpha_\mu^\textrm{FM}\equiv\alpha_\mu^\textrm{OP}$ is given by Eq.~(\ref{OPgaugefield}),\cite{ryder1980}
\begin{align}\label{QHFMgaugefield}
\alpha_\mu^\textrm{FM}(\b{r},t)=\frac{\partial n^I}{\partial x^\mu}\c{A}_I(\b{n}(\b{r},t))
=s(1-\cos\vartheta)\partial_\mu\varphi.
\end{align}
With this identification, our effective Lagrangian (\ref{LOP}) reads
\begin{align}\label{LagQHFM}
\c{L}=&-\frac{u}{2}(\rho-\bar{\rho})^2
-\rho(\partial_0\theta+\alpha_0^\textrm{FM}+\alpha_0+A_0)\nonumber\\
&-\frac{\kappa\rho}{2}(\partial_i\theta
+\alpha_i^\textrm{FM}+\alpha_i+A_i)^2
+\frac{1}{4\pi q}\epsilon^{\mu\nu\lambda}
\alpha_\mu\partial_\nu\alpha_\lambda\nonumber\\
&-\c{H}_\textrm{FM}(\b{n},\partial_i\b{n}),
\end{align}
where $\c{H}_\textrm{FM}$ is the Hamiltonian of the $O(3)$ NL$\sigma$M,\cite{Fradkin}
\begin{align}\label{HamO3NLsM}
\c{H}_\textrm{FM}=\frac{K}{2}(\partial_i\b{n})^2,\hspace{5mm}\b{n}^2=1,
\end{align}
and $K>0$ is the spin stiffness of the ferromagnet. The saddle-point analysis mentioned earlier can be repeated here, with $\b{n}$ constant and uniform in the ground state. Small fluctuations above the ground state are described by Eq.~(\ref{LagrangianQHOPdirect}), where the order parameter Lagrangian $\c{L}_\textrm{FM}\equiv\c{L}_\textrm{OP}$ is
\begin{align}\label{FMNLsigmaM}
\c{L}_\textrm{FM}=-\bar{\rho}\boldsymbol{\c{A}}(\b{n})\cdot
\partial_0\b{n}-\frac{K}{2}(\partial_i\b{n})^2,
\end{align}
a nonrelativistic version of the $O(3)$ NL$\sigma$M Lagrangian which is the low-energy effective action of a ferromagnet.\cite{Fradkin} The first term is a Berry phase term which is responsible for the $k^2$ dispersion of ferromagnetic magnons.

The dual theory is given by Eq.~(\ref{FQHOPMCS}), where the topological current $J^\mu_\textrm{FM}\equiv J^\mu_\textrm{OP}$ of the ferromagnet is\cite{wilczek1983}
\begin{align}\label{QHFMskyrmion}
J^\mu_\textrm{FM}=\frac{1}{2\pi}\epsilon^{\mu\nu\lambda}\partial_\nu\alpha_\lambda^\textrm{FM}
=\frac{1}{8\pi}\epsilon^{\mu\nu\lambda}\b{n}\cdot(\partial_\nu\b{n}\times\partial_\lambda\b{n}),
\end{align}
i.e., the skyrmion current.\cite{belavin1975} Integrating out the dual gauge field gives a long-range interaction between the skyrmions,\cite{sondhi1993}
\begin{align}
\delta S_\textrm{FM}=\int d^3x\int d^3x'\,
J^\mu_\textrm{FM}(x)\Pi_{\mu\nu}(x-x')
J^\nu_\textrm{FM}(x'),\nonumber
\end{align}
where the energetics and statistics of the skyrmions are determined by the symmetric and antisymmetric parts of the gauge field polarization $\Pi_{\mu\nu}$, respectively.

An alternative formulation of QH ferromagnetism\cite{sondhi1993} which will be useful for our discussion of nematic order is based on the $\mathbb{C}P^1$ formulation\cite{eichenherr1978,dadda1978,witten1979} of the $O(3)$ NL$\sigma$M. Equation (\ref{SO3SO2}) can be equivalently written as $\mathbb{S}^2=SU(2)/U(1)$.\cite{nakahara} A unit vector $\b{n}$ parameterizing $\mathbb{S}^2$ can be represented by a 2-component complex unit spinor $z_a=(z_1,z_2)\in\mathbb{C}^2$ with $|z_1|^2+|z_2|^2=1$ as
\begin{align}\label{HopfMap}
\b{n}=z_a^*\boldsymbol{\sigma}^{ab}z_b,
\end{align}
where $\boldsymbol{\sigma}=(\sigma_x,\sigma_y,\sigma_z)$ is the vector of Pauli matrices. Equation (\ref{HopfMap}), known as the first Hopf map,\cite{nakahara} is a map from the Lie group $SU(2)$ to $\mathbb{S}^2$. That is because $SU(2)$, the group of unitary $2\times 2$ matrices with unit determinant, can be parameterized by the set of matrices
\begin{align}\label{SU2def}
M=\left(\begin{array}{cc}
\alpha & \beta^* \\
-\beta & \alpha^*
\end{array}\right),\hspace{5mm}
\det M=|\alpha|^2+|\beta|^2=1,
\end{align}
with $(\alpha,\beta)\in\mathbb{C}^2$.
The Hopf map is however not one-to-one but many-to-one, since $\b{n}$ is invariant under the local $U(1)$ gauge transformation $z_a(x^\mu)\mapsto e^{i\theta(x^\mu)}z_a(x^\mu)$. Therefore, to represent $\b{n}$ faithfully we need to mod out these gauge transformations, and the target space $\mathbb{S}^2$ is obtained as the quotient space $SU(2)/U(1)=\mathbb{C}P^1\cong \mathbb{S}^2$. In terms of the $\mathbb{C}P^1$ field $z_a$, the order parameter gauge field (\ref{QHFMgaugefield}) is given by
\begin{align}\label{alphaOPCP1}
\alpha_\mu^\textrm{FM}=iz_a^*\partial_\mu z_a,
\end{align}
with $s=\frac{1}{2}$, as can be shown explicitly in the gauge $(z_1,z_2)=(\cos\frac{\vartheta}{2},e^{-i\varphi}\sin\frac{\vartheta}{2})$. The skyrmion current (\ref{QHFMskyrmion}) is given by
\begin{align}
J^\mu_\textrm{FM}=\frac{i}{2\pi}\epsilon^{\mu\nu\lambda}\partial_\nu(z_a^*\partial_\lambda z_a),\nonumber
\end{align}
and Eq.~(\ref{FMNLsigmaM}) becomes\cite{Fradkin}
\begin{align}\label{LCP1}
\c{L}_\textrm{FM}=-\bar{\rho}iz_a^*\partial_0z_a
-\frac{K}{2}\left(\partial_iz_a^*\partial_iz_a
+(z_a^*\partial_iz_a)^2\right).
\end{align}
Although the $\mathbb{C}P^1$ Lagrangian (\ref{LCP1}) appears to only have a global $U(1)$ symmetry $z_a\mapsto e^{i\theta}z_a$ with $\theta$ independent of $x^\mu$, it is in fact invariant under local $U(1)$ transformations with $\theta=\theta(x^\mu)$. Using the constraint $z_a^*z_a=1$, Eq.~(\ref{LCP1}) can be written in the manifestly gauge-invariant form
\begin{align}\label{LCP1b}
\c{L}_\textrm{FM}=-\bar{\rho}iz_a^*\partial_0z_a
-\frac{K}{2}|D_iz_a|^2,
\end{align}
with $D_i=\partial_i+i\alpha_i^\textrm{FM}$ the gauge-covariant derivative and $\alpha_i^\textrm{FM}$ given in Eq.~(\ref{alphaOPCP1}). Under a local $U(1)$ transformation, the Berry phase term in Eq.~(\ref{LCP1b}) changes by the total derivative $\bar{\rho}\partial_0\theta$. If we canonically quantize the theory (\ref{LCP1})-(\ref{LCP1b}), the Berry phase term implies the following choice of equal-time canonical commutation relations for the $\mathbb{C}P^1$ field $z_a$,
\begin{align}\label{schwinger}
[z_a(\b{r}),z_b^*(\b{r}')]=-2\pi q\ell_B^2\delta_{ab}\delta(\b{r}-\b{r}'),
\end{align}
i.e., the Schwinger boson algebra,\cite{Auerbach} which together with the Hopf map (\ref{HopfMap}) correctly reproduces the $\mathfrak{so}(3)$ algebra for $\b{n}$,
\begin{align}\label{SO3algebra}
[n^I(\b{r}),n^J(\b{r}')]=-4\pi q\ell_B^2i\epsilon^{IJK}n^K(\b{r})\delta(\b{r}-\b{r}').
\end{align}
As we have just seen, QH ferromagnetism is an example of the general procedure outlined at the beginning of Sec.~\ref{sec:OPinternal} with $M=\mathbb{S}^2$, $G=SU(2)$, and $H=U(1)$. As will be seen in the next section, the QH nematic problem studied in this paper is, in a precise sense to be defined [see Eq.~(\ref{analyticcont})], an analytic continuation of the QH ferromagnetism problem with $M=\mathbb{H}^2$ the hyperbolic plane or pseudosphere,\cite{balazs1986} $G=SU(1,1)$ the group of pseudounitary $2\times 2$ matrices, and $H=U(1)$.

\section{Spontaneous breaking of a spatial symmetry: nematic order}
\label{sec:QHN}

There are other symmetries, besides spin rotation symmetry, which can be spontaneously broken in the FQHE. In this paper we consider the possibility that the spatial $SO(2)$ rotation invariance is spontaneously broken, leading to the formation of a FQH nematic phase.\cite{NoteQHIN} We ignore the spin of the electron.

\subsection{Construction of a field theory}

The spontaneous breaking of $SO(2)$ rotation invariance is described by the Landau-de Gennes nematic order parameter $\hat{Q}_{ab}$, which is a real, symmetric, traceless $2\times 2$ matrix.\cite{DeGennes} The expectation value of $\hat{Q}_{ab}$ is zero in the isotropic phase and nonzero in the nematic phase. Instead of working with $\hat{Q}_{ab}$, we can equivalently consider the $2\times 2$ matrix $g_{ab}$ defined as the matrix exponential of $\hat{Q}_{ab}$,
\begin{align}\label{gexpQ}
g=\exp \hat{Q}.
\end{align}
Because $\hat{Q}_{ab}$ is real, symmetric, and traceless, $g_{ab}$ is real, symmetric, and unimodular ($\det g=1$). The most general matrix $g_{ab}$ of this form can be written as
\begin{align}\label{gparam}
g=\left(\begin{array}{cc}
T+X & Y \\
Y & T-X
\end{array}\right),\hspace{3mm}\det g=T^2-X^2-Y^2=1,
\end{align}
where $T,X,Y$ are real parameters. We see from Eq.~(\ref{gparam}) that $g_{ab}$, and by extension the nematic order parameter $\hat{Q}_{ab}$ from which it is constructed, corresponds to a unit Lorentz vector in 3D Minkowski space $\mathbb{R}^{2,1}$. The group of linear transformations on Minkowski space which preserve the Minkowski inner product (i.e., the length of Lorentz vectors) is the group of Lorentz transformations $SO^+(2,1)$.\cite{NoteLorentzSO21} The 2D surface in $\mathbb{R}^{2,1}$ parameterized by $\det g=1$ is a Riemannian real 2-manifold with constant negative Gaussian curvature known equivalently as the hyperbolic 2-space $\mathbb{H}^2$, the hyperbolic or Bolyai-Lobachevsky plane, or the pseudosphere.\cite{balazs1986} It is equivalent to the Euclidean plane $\mathbb{R}^2$ topologically but not geometrically, and can be viewed as an analytic continuation of the 2-sphere or elliptic 2-space $\mathbb{S}^2$. Unlike $\mathbb{S}^2$ however, the hyperbolic plane $\mathbb{H}^2$ cannot be embedded in 3D Euclidean space $\mathbb{R}^3$ because of its negative curvature, but it can be embedded in 3D Minkowski space $\mathbb{R}^{2,1}$. The target space for our nematic order parameter $\hat{Q}_{ab}$ is therefore the hyperbolic plane $\mathbb{H}^2$. However, as will be discussed in more detail later, this formulation includes in the target space not only the direction {{\it but also the amplitude of the nematic order}. Therefore, 
in contrast to the case of the QH ferromagnetism our effective field theory will be of the LGW type rather than of the NL$\sigma$M type.

\begin{widetext}
\begin{center}
\begin{table}
\begin{tabular}{|c||c|c|}
\hline
 & $SO(3)$ rotor & 2D nematic  \\
\hline
``classical'' degree of freedom & $\b{n}=(X,Y,Z)$ & $g=\left(\begin{array}{cc}T+X & Y \\ Y & T-X\end{array}\right)=\exp\hat{Q}$ \\
\hline
constraint & $\b{n}^2=1\Leftrightarrow X^2+Y^2+Z^2=1$ & $\det g=1\Leftrightarrow T^2-X^2-Y^2=1$ \\
\hline
corresponds to unit vector in... & 3D Euclidean space $\mathbb{R}^3$ & 3D Minkowski space $\mathbb{R}^{2,1}$ \\
\hline
direct linear isometry group & $SO(3)$ & $SO^+(2,1)$ \\
\hline
double cover & $SU(2)\cong\mathbb{S}^3$ & $SU(1,1)\cong AdS_3$ \\
\hline
``quantum'' degree of freedom & $SU(2)$ spinor $\left(\begin{array}{c}z_1 \\ z_2\end{array}\right)$ with $\delta^{ab}z_a^*z_b=1$
& $SU(1,1)$ spinor $\left(\begin{array}{c}z_1 \\ z_2\end{array}\right)$ with $i\epsilon^{ab}z_a^*z_b=1$ \\
\hline
$G\rightarrow G/H$ projection & $\b{n}=z_a^*\boldsymbol{\sigma}^{ab}z_b$ & $g_{ab}=z_a^*z_b+\mathrm{c.c.}$ \\
\hline
isotropy subgroup & $U(1)$ & $U(1)$ \\
\hline
target manifold & $SU(2)/U(1)=\mathbb{C}P^1=\mathbb{S}^2$: elliptic 2-space & $SU(1,1)/U(1)=\mathbb{H}^2$: hyperbolic 2-space \\
\hline
order parameter gauge field & $\alpha_\mu^\textrm{FM}=
iz_a^*\partial_\mu z_a$ & $\alpha_\mu^\textrm{N}=
-\epsilon^{ab}z_a^*\partial_\mu z_b+\ldots$ \\
\hline
\end{tabular}
\caption{2D nematic as the analytic continuation of a $SO(3)$ rotor.}
\label{table:nematic}
\end{table}
\end{center}
\end{widetext}

For simplicity, we follow an analog of the $\mathbb{C}P^1$ route of Sec.~\ref{sec:QHFM} to construct a gauge field from the nematic order parameter. The target space $M=\mathbb{H}^2$ is most naturally viewed as a quotient space $M=G/H$ where $G=SU(1,1)$ and $H=U(1)$. The Lie group $SU(1,1)$, defined as the group of complex $2\times 2$ matrices $M$ satisfying $M^\dag JM=J$ with $J=\diag(1,-1)$, can be parameterized in a way very similar to Eq.~(\ref{SU2def}) for $SU(2)$ by the set of matrices\cite{Gilmore}
\begin{align}\label{defSU11}
M=\left(\begin{array}{cc}
\alpha & \beta^* \\
\beta & \alpha^*
\end{array}\right),\hspace{5mm}
\det M=|\alpha|^2-|\beta|^2=1,
\end{align}
with $(\alpha,\beta)\in\mathbb{C}^2$. Just as the unit vector $\b{n}$ parameterizing $\mathbb{S}^2$ can be represented by a 2-component complex field $z_a$ subject to the constraint $z_a^*z_a=1$ [Eq.~(\ref{HopfMap})], so the unimodular metric $g_{ab}$ parameterizing $\mathbb{H}^2$ can be represented by a 2-component complex field $z_a=(z_1,z_2)\in\mathbb{C}^2$ as
\begin{align}\label{defgabz}
g_{ab}=z_a^*z_b+z_b^*z_a,
\end{align}
where the $z_a$ satisfy the constraint
\begin{align}\label{constraintz}
i\epsilon^{ab}z_a^*z_b=1.
\end{align}
Similar to the Hopf map $SU(2)\rightarrow \mathbb{S}^2$ [Eq.~(\ref{HopfMap})], Eq.~(\ref{defgabz}) can be viewed as a map $SU(1,1)\rightarrow \mathbb{H}^2$. Indeed, choosing $\alpha=\frac{1}{\sqrt{2}}(z_1+iz_2)$ and $\beta=\frac{1}{\sqrt{2}}(z_1-iz_2)$, we have $|\alpha|^2-|\beta|^2=i\epsilon^{ab}z_a^*z_b$, hence the space defined by the constraint (\ref{constraintz}) is equivalent to $SU(1,1)$ [see Eq.~(\ref{defSU11})]. As in the case of the Hopf map, the map $SU(1,1)\rightarrow \mathbb{H}^2$ has a $U(1)$ redundancy since $g_{ab}$ is invariant under local $U(1)$ gauge transformations $z_a(x^\mu)\mapsto e^{i\theta(x^\mu)}z_a(x^\mu)$. To represent $g_{ab}$ faithfully we need to mod out these gauge transformations, and the target space $\mathbb{H}^2$ is obtained\cite{Gilmore} as the quotient space $SU(1,1)/U(1)=\mathbb{H}^2$. Table~\ref{table:nematic} summarizes the analogies between the $SO(3)$ rotor order parameter of QH ferromagnetism and our $SO^+(2,1)$ ``Minkowski'' order parameter for QH nematics.

We digress to note that Eq.~(\ref{defgabz}) is also Haldane's expression for the unimodular intrinsic metric of the FQHE in terms of what he calls the ``geometry field'' $z_a(\b{r},t)$.\cite{haldane2011} Indeed, the $z_a$ have a natural interpretation in terms of the tetrads or ``vielbeins'' (zweibeins in 2D) of Riemannian geometry.\cite{nakahara} The zweibeins of a 2D Riemannian manifold are two real, non-coordinate basis vectors $\b{e}_1=(e_1^x,e_1^y)$ and $\b{e}_2=(e_2^x,e_2^y)$ which encode much of the geometry of the manifold. In particular, the metric tensor is given by $g_{ab}=\b{e}_a\cdot\b{e}_b$, such that the zweibeins can be viewed as the ``square root'' of the metric. The ``geometry field'' $z_a$ is 
nothing else but a complex linear combination of the zweibeins, $z_a=\frac{1}{\sqrt{2}}(e_a^x-ie_a^y)$, $a=1,2$. The unimodular constraint (\ref{constraintz}) is equivalent to $\b{e}_1\wedge\b{e}_2=1$, where $\b{e}_1\wedge\b{e}_2$ is the area of the parallelogram spanned by the basis vectors $\b{e}_a$. Because of the group isomorphism $SO^+(2,1)\cong PSL(2,\mathbb{R})=SL(2,\mathbb{R})/\mathbb{Z}_2$,\cite{witten1988} a $SO^+(2,1)$ Lorentz transformation of the Minkowski 3-vector $(T,X,Y)$ parameterizing $g_{ab}$ [see Eq.~(\ref{gparam})] is equivalent to a change of basis $\b{e}_a\mapsto\b{e}'_a=P_{aa'}\b{e}_{a'}$ with $P\in PSL(2,\mathbb{R})$. Because such a change of basis preserves the area $\b{e}_1'\wedge\b{e}_2'=\b{e}_1\wedge\b{e}_2=1$, the $SO^+(2,1)$ ``rotations'' of the metric $g_{ab}$ correspond in this sense to linear area-preserving diffeomorphisms.\cite{HaldaneMetric,haldane2011} From now on, we refer to $g_{ab}$ as the metric and to $z_a$ as the zweibein, and return to our program of constructing a field theory of the isotropic-to-nematic transition.

As explained in Sec.~\ref{sec:OPinternal}, we need to construct a $U(1)$ gauge field $\alpha_\mu^\textrm{N}$ from the nematic degrees of freedom. The analogy between the $\mathbb{C}P^1=SU(2)/U(1)$ formulation of QH ferromagnetism and the $SU(1,1)/U(1)$ description of 2D nematics suggests the following definition of $\alpha_\mu^\textrm{N}$,
\begin{align}\label{alphaOPnematicsz}
\alpha_\mu^\textrm{N}=-\epsilon^{ab}z_a^*\partial_\mu z_b+\ldots
\end{align}
Indeed, the first term has the same transformation property $\alpha_\mu^\textrm{N}\mapsto\alpha_\mu^\textrm{N}-\partial_\mu\theta$ under a $U(1)$ gauge transformation $z_a\mapsto e^{i\theta}z_a$ as the $\mathbb{C}P^1$ gauge field (\ref{alphaOPCP1}), due to the constraint (\ref{constraintz}). The remaining terms ($\ldots$) should therefore be gauge invariant and will be discussed shortly. We can obtain an explicit expression for the first term in Eq.~(\ref{alphaOPnematicsz}) which is similar to Eq.~(\ref{QHFMgaugefield}) for the ferromagnetic gauge field by adopting the following parameterization of the target space $\mathbb{H}^2$,
\begin{align}
z_1&=\frac{1}{\sqrt{2}}\left(\cosh\textstyle\frac{Q}{2}+e^{-i\varphi}\sinh\textstyle\frac{Q}{2}\right),\label{z1}\\
z_2&=-\frac{i}{\sqrt{2}}\left(\cosh\textstyle\frac{Q}{2}-e^{-i\varphi}\sinh\textstyle\frac{Q}{2}\right),\label{z2}
\end{align}
modulo $U(1)$ gauge transformations $z_a\mapsto e^{i\theta}z_a$. The pseudospherical coordinates\cite{balazs1986} $Q\geq 0$ and $0\leq\varphi<2\pi$ on $\mathbb{H}^2$ are 
nothing but the amplitude and phase of the Landau-de Gennes nematic order parameter $\hat{Q}_{ab}$,
\begin{align}\label{Qcmplx}
\hat{Q}_{11}+i\hat{Q}_{12}=Qe^{i\varphi},
\end{align}
or, equivalently,
\begin{align}\label{Qab}
\hat{Q}_{ab}=Q(2\hat{d}_a\hat{d}_b-\delta_{ab}),
\end{align}
where
\begin{align}\label{director}
\hat{\b{d}}=(\cos\textstyle\frac{\varphi}{2},\sin\frac{\varphi}{2}),
\end{align}
is the nematic director,\cite{DeGennes} as can be checked explicitly by using Eq.~(\ref{z1}), (\ref{z2}), (\ref{defgabz}), and (\ref{gexpQ}). We note that $\hat{\b{d}}\rightarrow-\hat{\b{d}}$  under $\varphi\rightarrow\varphi+2\pi$, i.e., $\varphi$ is already defined in such a way that all values of $\varphi$ between zero and $2\pi$ correspond to physically distinct states, even when we take into account the fact that the nematic order parameter is a headless vector. This will make the notion of a $2\pi$ vortex more comfortable when we discuss topological defects in the nematic phase later on. Substituting Eq.~(\ref{z1})-(\ref{z2}) into Eq.~(\ref{alphaOPnematicsz}), we obtain
\begin{align}\label{NematicGaugeFieldExplicit}
\alpha_\mu^\textrm{N}=\half(1-\cosh Q)\partial_\mu\varphi+\ldots,
\end{align}
which can be viewed as an analytic continuation of the ferromagnetic gauge field (\ref{QHFMgaugefield}),
\begin{align}\label{analyticcont}
\alpha_\mu^\textrm{N}(Q,\varphi)=\alpha_\mu^\textrm{FM}(\vartheta\rightarrow iQ,\varphi),
\end{align}
modulo the gauge invariant terms ($\ldots$). A 2D nematic can thus be viewed as an $SO(3)$ rotor with imaginary polar angle, and an $SO(3)$ rotor can be viewed as a 2D nematic with imaginary amplitude.

We now discuss the gauge invariant terms ($\dots$) in Eq.~(\ref{alphaOPnematicsz}). 
In contrast with QH ferromagnetism, nematic order corresponds to the breaking of a spatial symmetry, $SO(2)$ rotation symmetry. The zweibein $z_a$ transforms as a vector $z_a\rightarrow R_{aa'}z_{a'}$ under spatial rotations $R\in SO(2)$. Likewise, the spatial part of the nematic gauge field $\alpha^\textrm{N}_c$ with $c$ a spatial index transforms as a vector. In the notation of Eq.~(\ref{OPgaugefield}), for an internal symmetry the order parameter indices $I$ are contracted independently of the spacetime index $\mu$, but for a spatial symmetry $I$ and $\mu$ can be contracted together when $\mu=c$ is a spatial index. For $\mu=0$, this cannot happen and we have $\alpha_0^\textrm{N}=-\epsilon^{ab}z_a^*\partial_0z_b$ with no extra terms. We now investigate the possible extra terms in $\alpha_c^\textrm{N}$. What we are really after is a description of the vicinity of the isotropic-to-nematic transition where the amplitude $Q$ of the nematic order is small. We should therefore expand Eq.~(\ref{alphaOPnematicsz}) in powers of $Q$. Furthermore, we are interested in the long-wavelength limit, and should also expand in powers of spatial derivatives. We find
\begin{align}\label{alphaNc}
\alpha_c^\textrm{N}=
-\frac{1}{8}\epsilon_{bd}\hat{Q}_{ab}\partial_c \hat{Q}_{da}+\ldots
\end{align}
However, we can clearly write down two lower order terms, $\partial_a\hat{Q}_{ac}$ and $\epsilon_{ab}\partial_a\hat{Q}_{bc}$, which are consistent with rotation symmetry. Whether to keep one or both terms cannot be decided on symmetry grounds alone and we invoke a physical argument. From Eq.~(\ref{OPtopcurrent}), the density of topological excitations of the order parameter is given by the flux of the order parameter gauge field $J^0_\textrm{N}=\frac{1}{2\pi}\epsilon_{ab}\partial_a\alpha_b^\textrm{N}$. In the QH ground state [Eq.~(\ref{FQHOPMCS})], the electric charge density $J^0$ is equal to the flux of the dual gauge field $\b{a}$ which, in the absence of Laughlin quasiparticles, is equal to $\nu$ times $J^0_\textrm{N}$. On the one hand, the contribution of the first term in Eq.~(\ref{alphaNc}) to $J^0$ corresponds to the electric charge carried by nematic disclinations as will be seen later. On the other hand, 
on symmetry grounds it is natural to identify the nematic order parameter $\hat{Q}_{ab}$ with the local electric quadrupole moment.
 A straightforward extension of the standard coarse-graining procedure\cite{jackson} used to derive the macroscopic Maxwell's equations shows that in 2D, a spatially varying electric quadrupole moment $\hat{Q}_{ab}$ induces a charge density $\delta J^0=\frac{1}{4}\partial_a\partial_b\hat{Q}_{ab}$, which corresponds to adding a term $\propto\epsilon_{ab}\partial_a\hat{Q}_{bc}$ in $\alpha^\textrm{N}_c$. Our final expression for the spatial part of the nematic gauge field is thus
\begin{align}\label{alphaNcfinal}
\alpha^\textrm{N}_c=-\frac{1}{8}\epsilon_{bd}\hat{Q}_{ab}\partial_c\hat{Q}_{da}
+C\epsilon_{ab}\partial_a\hat{Q}_{bc},
\end{align}
where $C$ is a constant to be determined. As will be seen later, $C$ is related to the coefficient of the $k^4$ term in the equal-time structure factor $S(\b{k})=\langle[J^0(\b{k}),J^0(-\b{k})]\rangle$.

\begin{figure}[t]
\begin{center}
\includegraphics[width=\columnwidth]{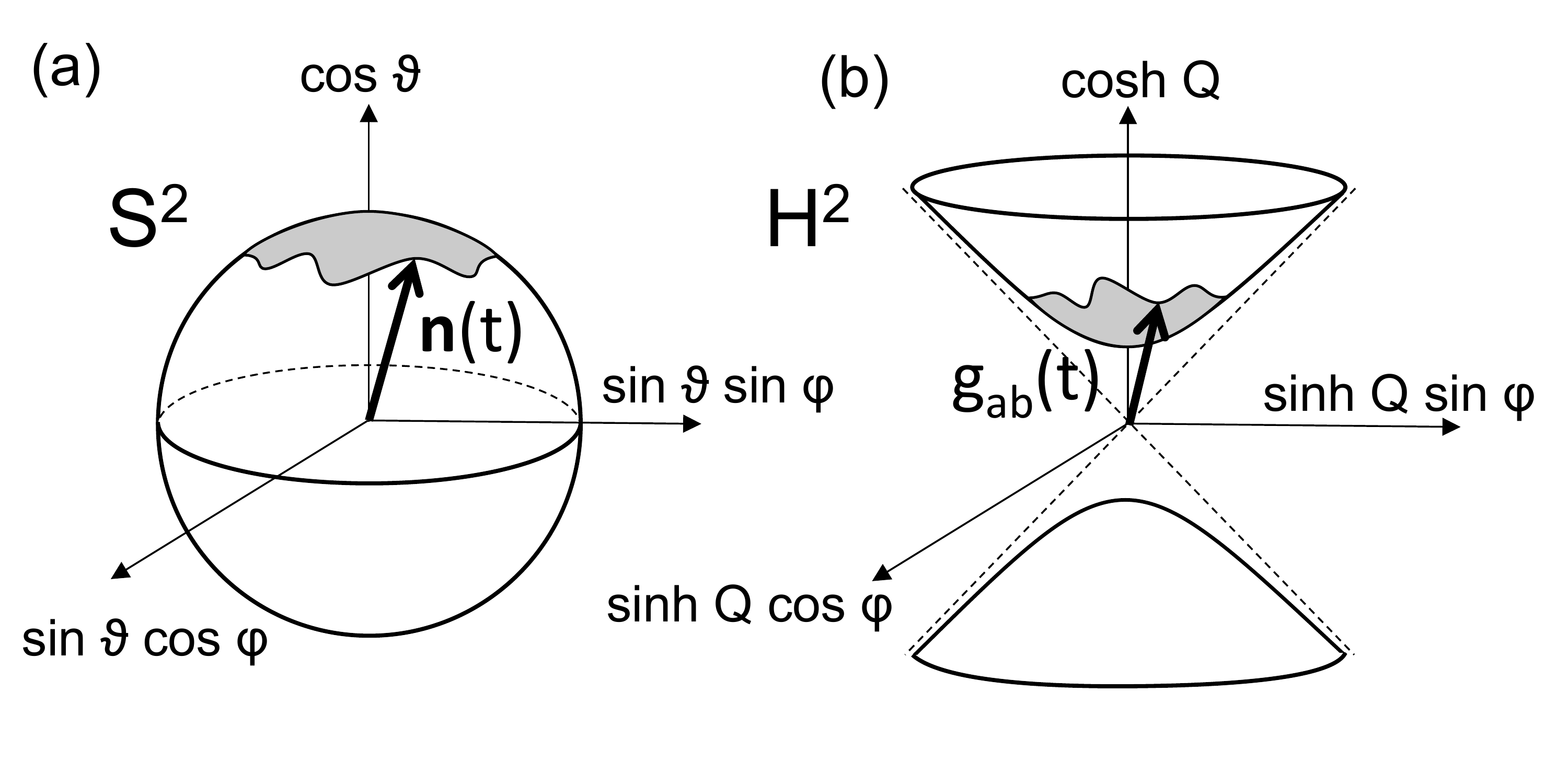}
\end{center}
\caption{The time component $\alpha_0^\textrm{OP}$ of the order parameter gauge field is a Berry phase term which measures the area enclosed by a trajectory of the order parameter on the target space $M$. (a) For the ferromagnet, $M=\mathbb{S}^2$; (b) for the nematic, $M=\mathbb{H}^2$.}
\label{fig:berry}
\end{figure}

As seen in Sec.~\ref{sec:QHFM}, the time component $\alpha_0^\textrm{FM}=iz_a^*\partial_0z_a$ of the ferromagnetic gauge field has the interpretation of a Wess-Zumino or Berry phase term for the ferromagnetic order parameter $\b{n}$ [Fig.~\ref{fig:berry}(a)]. The time integral $\int dt\,\alpha_0^\textrm{FM}$ is equal to one half of the area $\mathbb{A}[\b{n}(t)]$ (modulo $4\pi$) of the surface on $\mathbb{S}^2$ bounded by the curve $\b{n}(t)$.\cite{Fradkin} Likewise, in the nematic case we have
\begin{align}\label{BerryPhaseNematic}
\int dt\,\alpha_0^\textrm{N}&=\half\int(1-\cosh Q)\dot{\varphi}\,dt\nonumber\\
&=\half\int_0^{2\pi}(1-\cosh Q(\varphi))\,d\varphi\nonumber\\
&=-\half\int_0^{2\pi}d\varphi\int_1^{\cosh Q(\varphi)}d(\cosh Q)\nonumber\\
&=-\half\int_0^{2\pi}\int_0^{Q(\varphi)}\sinh Q\,dQ\,d\varphi\nonumber\\
&=-\half\mathbb{A}[g_{ab}(t)],
\end{align}
where $\mathbb{A}[g_{ab}(t)]$ is the area of the surface on $\mathbb{H}^2$ bounded by the curve $g_{ab}(t)$ [Fig.~\ref{fig:berry}(b)]. Indeed, the metric $ds_{\mathbb{H}^2}^2$ on $\mathbb{H}^2$ (see Ref.~\onlinecite{balazs1986}) can be obtained from the usual metric $ds_{\mathbb{S}^2}^2=d\vartheta^2+\sin^2\vartheta\,d\varphi^2$ on $\mathbb{S}^2$ by analytic continuation $\vartheta\rightarrow iQ$,
\begin{align}\label{metricH2}
ds_{\mathbb{H}^2}^2=-(dQ^2+\sinh^2Q\,d\varphi^2)=-G_{IJ}(\mathbb{H}^2)d\Phi^Id\Phi^J,
\end{align}
where $\Phi^I=(Q,\varphi)$ and $G_{IJ}(\mathbb{H}^2)=\diag(1,\sinh^2Q)$, from which we obtain the volume form on $\mathbb{H}^2$,
\begin{align}
\mathrm{vol}(\mathbb{H}^2)=\sqrt{\det G(\mathbb{H}^2)}\,d\Phi^1\wedge d\Phi^2=\sinh Q\,dQ\wedge d\varphi,\nonumber
\end{align}
and $\mathbb{A}=\int\mathrm{vol}(\mathbb{H}^2)$, which gives us Eq.~(\ref{BerryPhaseNematic}).

Equations~(\ref{alphaOPnematicsz}) and (\ref{alphaNcfinal}) are closely related to another important quantity in Riemannian geometry besides the metric tensor which can be constructed out of the zweibeins: the spin connection.\cite{nakahara} For a 2D Riemannian manifold, the spin connection $\Omega_c{}^A{}_B$ is a 1-form valued in the $\mathfrak{so}(2)$ Lie algebra, with $A,B$ the non-coordinate $\mathfrak{so}(2)$ indices and $c$ the coordinate (1-form) index. It is given in terms of the real zweibeins $e_a^A$ by $\Omega_c{}^A{}_B=-e^a_B\nabla_ce^A_a=\epsilon^A{}_B\Omega_c$ where the inverse zweibeins $e_B^b$ satisfy $e_b^Ae^b_B=\delta^A_B$, and $\nabla$ is the covariant derivative, e.g., $\nabla_cX_a=\partial_cX_a-\Gamma^b_{ca}X_b$ with $\Gamma^b_{ca}$ the connection coefficients. In terms of the complex zweibeins $z_a$, we have
\begin{align}
\Omega_c=-\epsilon^{ab}z_a^*\partial_c z_b-i\Gamma^e_{cb}z^{b*}z_e.\nonumber
\end{align}
The first (gauge variant) term in $\Omega_c$ corresponds to the first term in Eq.~(\ref{alphaNcfinal}), whereas the gauge invariant term involving the Christoffel symbols $\Gamma^e_{cb}$, when expanded to leading order in $Q$, reproduces the second term of Eq.~(\ref{alphaNcfinal}) with $C=-\frac{1}{2}$. In our effective field theory, we simply consider $C$ as a parameter.

Before coupling the nematic degrees of freedom to the CS gauge field in either the direct or dual approaches to the FQHE, we need to construct a Lagrangian $\c{L}_\textrm{N}$ for the nematic degrees of freedom alone, i.e., the nematic analog of $\c{L}_\textrm{FM}$. If written in terms of the zweibeins $z_a$, $\c{L}_\textrm{N}$ should be invariant under local $U(1)$ gauge transformations $z_a\mapsto e^{i\theta}z_a$, because $\hat{Q}_{ab}$ is invariant under such transformations. In analogy to the $\mathbb{C}P^1$ Lagrangian (\ref{LCP1})-(\ref{LCP1b}), we 
write
\begin{align}\label{LN}
\c{L}_\textrm{N}=\lambda\epsilon^{ab}z_a^*\partial_0z_b-\c{H}_\textrm{N},
\end{align}
where $\c{H}_\textrm{N}$ is a Hamiltonian to be determined below, and the first term is a Berry phase term which, as explained earlier in the case of the ferromagnet, is generally allowed in the long-wavelength effective Lagrangian of a time-reversal symmetry breaking state such as the FQH state. In Sec.~\ref{sec:IsotropicPhase} [see Eq.~(\ref{lambda})] we argue that $\lambda=\bar{s}\bar{\rho}$ where $\bar{s}$ is the mean orbital spin per particle\cite{read2009,read2011} or guiding-center spin,\cite{HaldaneMetric,haldane2011,HaldaneFieldTheory} a quantized property of time-reversal symmetry breaking topological phases. Upon quantization, Eq.~(\ref{LN}) imposes the following canonical commutation relations for the zweibeins,\cite{HaldaneFieldTheory}
\begin{align}\label{zweibeincomm}
[z_a(\b{r}),z_b^*(\b{r}')]=-2\pi q\ell_B^2\bar{s}^{-1}i\epsilon_{ab}\delta(\b{r}-\b{r}').
\end{align}
Just as the choice (\ref{schwinger}) of commutation relations for the $\mathbb{C}P^1$ field gives rise to the $\mathfrak{so}(3)$ algebra (\ref{SO3algebra}) for the ferromagnetic order parameter $\b{n}$, the choice (\ref{zweibeincomm}) of commutation relations for the zweibeins gives rise to the $\mathfrak{so}(2,1)$ algebra for the metric $g_{ab}$,\cite{HaldaneMetric,haldane2011}
\begin{align}
[g_{ab}(\b{r}),g_{cd}(\b{r}')]=&-2\pi q\ell_B^2\bar{s}^{-1} i\bigl(\epsilon_{bc}g_{ad}(\b{r})+\epsilon_{ad}g_{bc}(\b{r})\nonumber\\
&+\epsilon_{bd}g_{ac}(\b{r})+\epsilon_{ac}g_{bd}(\b{r})\bigr)\delta(\b{r}-\b{r}').\nonumber
\end{align}

Pursuing our analogy with ferromagnetism, the Hamiltonian $\c{H}_\textrm{N}$ should be a $SO^+(2,1)$ analog of the $SO(3)$ NL$\sigma$M Hamiltonian (\ref{HamO3NLsM}). Using the expression of $\b{n}$ in spherical coordinates, we have $\c{H}_\textrm{FM}=\frac{K}{2}\left[(\partial_i\vartheta)^2+\sin^2\vartheta(\partial_i\varphi)^2\right]$, which can be written in the geometrical form $\c{H}_\textrm{FM}=\frac{K}{2}G_{IJ}(\mathbb{S}^2)\partial_i\Phi^I\partial_i\Phi^J$ where $\Phi^I=(\vartheta,\varphi)$ and $G_{IJ}(\mathbb{S}^2)=\diag(1,\sin^2\vartheta)$ is the metric tensor on $\mathbb{S}^2$. This suggests the choice $\c{H}_\textrm{N}=\frac{\kappa_Q}{4} G_{IJ}(\mathbb{H}^2)\partial_i\Phi^I\partial_i\Phi^J=\frac{\kappa_Q}{4}\left[(\partial_iQ)^2+\sinh^2Q(\partial_i\varphi)^2\right]$, where $G_{IJ}(\mathbb{H}^2)$ is the metric tensor on $\mathbb{H}^2$ [Eq.~(\ref{metricH2})] and $\kappa_Q$ is a nematic stiffness parameter (the factor of $\frac{1}{4}$ is conventional). As we briefly alluded to earlier, the target space $\mathbb{H}^2$ contains not only the direction $\varphi$ of the nematic order parameter $\hat{Q}_{ab}$, which is a compact variable, but also its amplitude $Q$, which is noncompact. While the ferromagnetic $\sigma$ model (\ref{FMNLsigmaM}) has a compact global $SO(3)$ symmetry which rotates $\b{n}$, the nematic $\sigma$ model (\ref{LN}) has a noncompact global $SO^+(2,1)$ symmetry which comprises not only $SO(2)$ rotations of the director $\hat{\b{d}}$, but also ``boosts'' of the nematic amplitude $Q$. In other words, $\c{H}_\textrm{N}$ so far describes a system for which all nematic amplitudes $Q\geq 0$ are equally energetically favorable, which is evidently unphysical.

In order to describe a physical isotropic-to-nematic transition where $\langle Q\rangle=0$ in the isotropic phase and $\langle Q\rangle\neq 0$ in the nematic phase, we need to add to $\c{H}_\textrm{N}$ a potential energy term $V_\textrm{N}$ which explicitly breaks the noncompact $SO^+(2,1)$ symmetry to $SO(2)$ by favoring a certain finite value of $Q$. A natural choice from a field theory standpoint is a $\phi^4$-type potential,
\begin{align}
V_\textrm{N}(Q,\varphi)=\frac{u_1}{2}(|z_a|^2-g)^2=\frac{u_1}{2}(\cosh Q-g)^2,\nonumber
\end{align}
with $u_1>0$, which is minimized by $Q=0$ for $g<1$ and by $Q=\cosh^{-1}g\neq 0$ for $g>1$, i.e., this potential has a continuous isotropic-to-nematic transition at $g=1$. We therefore consider the $SO(2)$-invariant Lagrangian
\begin{align}\label{LagN}
\c{L}_\textrm{N}=-\half\bar{s}\bar{\rho}(1-\cosh Q)\partial_0\varphi-\c{H}_\textrm{N},
\end{align}
where the nematic Hamiltonian is
\begin{align}
\c{H}_\textrm{N}=\frac{\kappa_Q}{4}\left[(\partial_iQ)^2+\sinh^2Q(\partial_i\varphi)^2\right]+\frac{u_1}{2}(\cosh Q-g)^2.\nonumber
\end{align}
Close to the transition, we have $Q\ll 1$ and Eq.~(\ref{LagN}) reads
\begin{align}\label{LagQmatrix}
\c{L}_\textrm{N}=\frac{\bar{s}\bar{\rho}}{8}\epsilon_{bc}
\hat{Q}_{ab}\partial_0\hat{Q}_{ca}
-\frac{\kappa_Q}{8}(\partial_c\hat{Q}_{ab})^2
-\frac{\tilde{r}}{2}\hat{Q}_{ab}^2-\frac{\tilde{u}}{4}(\hat{Q}_{ab}^2)^2,
\end{align}
where $\tilde{r}\propto(1-g)$ and $\tilde{u}>0$ are constants. We note that there are no cubic terms $\propto Q^3$ in the Lagrangians (\ref{LagN})-(\ref{LagQmatrix}), which reflects the vanishing in 2D of all terms $\tr(\hat{Q}^n)=0$, $n$ odd.

We are now in a position to construct a field theory of the FQH isotropic-to-nematic transition following Eq.~(\ref{LOP}). We have
\begin{align}\label{LagQHN}
\c{L}=&-\frac{u}{2}(\rho-\bar{\rho})^2-\rho(\partial_0\theta+\alpha_0^\textrm{N}+\alpha_0+A_0)\nonumber\\
&-\frac{\kappa\rho}{2}(\partial_i\theta+\alpha_i^\textrm{N}+\alpha_i+A_i)^2
+\frac{1}{4\pi q}\epsilon^{\mu\nu\lambda}
\alpha_\mu\partial_\nu\alpha_\lambda\nonumber\\
&-\c{H}_\textrm{N},
\end{align}
which is our central result. 

The reader will note that Eq.~(\ref{LagQHN}) parallels closely the Lagrangian proposed by Haldane\cite{HaldaneFieldTheory} with a few
caveats that flow from our very different conceptual
framework. 
First, the spirit of our LGW construction dictates
that the nematic order parameter $\hat{Q}_{ab}\ll 1$ whence Haldane's
metric $g_{ab}$ must be close to the background metric. Second, we are not
led to any deeper meaning for the geometrical constructs, such as covariant
derivatives, that are prominent in Haldane's thinking. In our analysis
the different pieces of the ``covariant spin connection'' arise from distinct 
pieces of physics. Third, we keep explicit track of a $U(1)$ field $\theta(\b{r},t)$ in order
to introduce Laughlin quasiparticles into the theory---they will {\it not}
arise as defects or topological excitations of the nematic/geometry field.
Fourth, we are led naturally to include a term quadratic in the current which 
Haldane excludes. This will reintroduce the (high-energy) Kohn mode into the theory which is pushed to infinite energy in its absence.

The ground state corresponds to the mean-field solution Eq.~(\ref{FQHOPsolution}) with a constant value of the nematic order parameter $\hat{Q}_{ab}$ given by
\begin{align}\label{MFFQHNsolQ}
\hat{Q}_{11}+i\hat{Q}_{12}=\left\{
\begin{array}{cc}
0, & g<1\text{ (isotropic phase)}, \\
\bar{Q}e^{i\bar{\varphi}}, & g>1\text{ (nematic phase)},
\end{array}
\right.
\end{align}
where $\bar{Q}=\cosh^{-1}g\sim\sqrt{2(g-1)}$ and $\bar{\varphi}$ is picked by spontaneous symmetry breaking. The Lagrangian for the Gaussian fluctuations above the mean-field ground state is given by Eq.~(\ref{LagrangianQHOPdirect}),
\begin{align}\label{LagFluctQHN}
\c{L}=&-\frac{u}{2}(\delta\rho)^2-\delta\rho(\partial_0\theta+\alpha_0^\textrm{N}+\delta\alpha_0)
\nonumber\\
&-\frac{\kappa\bar{\rho}}{2}(\partial_i\theta+\alpha_i^\textrm{N}+\delta\alpha_i)^2
+\frac{1}{4\pi q}\epsilon^{\mu\nu\lambda}
\delta\alpha_\mu\partial_\nu\delta\alpha_\lambda
\nonumber\\
&+\c{L}_\textrm{N},
\end{align}
where the nematic Lagrangian (\ref{LagN}) is expressed in terms of the zweibeins as
\begin{eqnarray}\label{LNzweibein1}
\c{L}_\textrm{N}&=&\bar{s}\bar{\rho}\epsilon^{ab}z_a^*\partial_0z_b
+\kappa_Q\left(i\epsilon^{ab}\partial_iz_a^*\partial_iz_b+(i\epsilon^{ab}z_a^*\partial_iz_b)^2\right)\nonumber\\
&&-\frac{u_1}{2}(|z_a|^2-g)^2.
\end{eqnarray}
The particle-vortex dual of Eq.~(\ref{LagFluctQHN}) is given by Eq.~(\ref{FQHOPMCS}),
\begin{align}\label{LdualFQHN}
\c{L}_\textrm{dual}=&\frac{q}{4\pi}\epsilon^{\mu\nu\lambda}a_\mu\partial_\nu a_\lambda
-(J^\mu_\textrm{vor}+J^\mu_\textrm{N})a_\mu\nonumber\\
&-\frac{1}{2\pi}\epsilon^{\mu\nu\lambda}A_\mu\partial_\nu a_\lambda+\c{L}_\textrm{N},
\end{align}
where $J^\mu_\textrm{vor}$ is the same as in Sec.~\ref{sec:QHOP}, and the topological current $J^\mu_\textrm{N}\equiv J^\mu_\textrm{OP}$ is
\begin{align}\label{JtopologicalNematic}
J^\mu_\textrm{N}=\frac{1}{2\pi}\epsilon^{\mu\nu\lambda}
\partial_\nu\alpha_\lambda^\textrm{N}.
\end{align}

\subsection{Isotropic phase}
\label{sec:IsotropicPhase}

The isotropic phase $g<1$ corresponds to an ordinary rotationally invariant FQH state. The Hall conductivity $\sigma_{xy}$ is quantized to $(1/q)(e^2/h)$, as follows from the topological part (\ref{Ltop}) of our Lagrangian, and the electrically neutral nematic fluctuations do not affect this result. The coefficient $\lambda$ in Eq.~(\ref{LN}) is fixed by the Hall viscosity as we now explain. In the same way that the uniform dc conductivity tensor $\sigma_{ab}$ is obtained by varying the background electromagnetic field $A_\mu\rightarrow A_\mu+\delta A_\mu$ and computing the linear response function
\begin{align}
\sigma_{ab}=\lim_{\omega\rightarrow 0}\frac{1}{i\omega}
\frac{\delta^2S_\textrm{eff}[\delta A_\mu]}{\delta A_a(-\omega)\delta A_b(\omega)},\nonumber
\end{align}
where $S_\mathrm{eff}[\delta A_\mu]$ is the effective action obtained by integrating out all dynamical fields, the uniform dc viscosity tensor $\eta_{abcd}$ is obtained by varying the background metric $\bar{g}_{ab}\rightarrow\bar{g}_{ab}+\delta\bar{g}_{ab}$ and computing\cite{avron1995,read2009,read2011,bradlyn2012}
\begin{align}
\eta_{abcd}=\lim_{\omega\rightarrow 0}\frac{1}{i\omega}
\frac{\delta^2S_\textrm{eff}[\delta\bar{g}_{ab}]}{\delta\bar{g}_{ab}(-\omega)\delta\bar{g}_{cd}(\omega)},\nonumber
\end{align}
where once again, $S_\mathrm{eff}[\delta\bar{g}_{ab}]$ is the effective action with all dynamical fields integrated out. We denote the background metric by $\bar{g}_{ab}$ to differentiate it from the dynamical metric $g_{ab}$. The Hall viscosity tensor $\eta^A_{abcd}=-\eta^A_{cdab}$ is the antisymmetric part of $\eta_{abcd}$. For an isotropic fluid on the 2D plane the Hall viscosity tensor has a single nonzero independent component\cite{avron1995} $\eta^A_{1112}=\eta^A_{1222}=\eta_H$.

We now show that in the isotropic phase $g<1$, the field theory (\ref{LagQmatrix}) gives the correct Hall viscosity predicted\cite{read2009} for the $\nu=1/q$ Laughlin FQH state. In the isotropic phase, we can neglect the quartic term $(\hat{Q}_{ab}^2)^2$ in Eq.~(\ref{LagQmatrix}). So far our theory is formulated for a flat background metric $\bar{g}_{ab}=\delta_{ab}$, and we need to specify how the nematic order parameter $\hat{Q}_{ab}$ couples to a deformed background metric $\delta_{ab}+\delta\bar{g}_{ab}$. For small nematic deformations, the quadratic term $\hat{Q}_{ab}^2$ is essentially $(g_{ab}-\delta_{ab})^2$ where $g_{ab}$ is the dynamical metric (\ref{defgabz}). A natural way of introducing a deformed background metric is to replace in this expression $\delta_{ab}$ by the deformed metric $\bar{g}_{ab}=\delta_{ab}+\delta\bar{g}_{ab}$. This is in agreement with the ideas of Ref.~\onlinecite{haldane2011,HaldaneFieldTheory} according to which for isotropic Coulomb interactions, there is an energy cost quadratic in the deviation of the dynamical metric $g_{ab}$ from its mean-field value which is set by the geometry of the effective mass tensor, i.e., the background metric. Expanding $g_{ab}$ as $g_{ab}\simeq\delta_{ab}+\hat{Q}_{ab}$ which is valid close to the transition, we obtain the Lagrangian
\begin{align}\label{CollModesIsotropic}
\c{L}_\textrm{N}=-\frac{i\bar{s}\bar{\rho}}{4}\c{Q}^*\partial_0{\c{Q}}-\frac{\kappa_Q}{4}|\partial_i\c{Q}|^2-\frac{\tilde{r}}{2}(\hat{Q}_{ab}-\delta\bar{g}_{ab})^2,
\end{align}
where $\c{Q}=\hat{Q}_{11}+i\hat{Q}_{12}=Qe^{i\varphi}$ [Eq.~(\ref{Qcmplx})]. Upon integrating out the dynamical fields $\c{Q}$, $\c{Q}^*$, we obtain the effective Lagrangian
\begin{align}
\c{L}_\textrm{eff}[\delta\bar{g}_{ab}]=
\frac{\bar{s}\bar{\rho}}{8}\epsilon^{ab}
\delta\bar{g}_{ac}\partial_0\delta\bar{g}_{bc}+\c{O}(\partial_0^2),\nonumber
\end{align}
in agreement with Eq.~(30) of Ref.~\onlinecite{hoyos2012}, which corresponds to a Hall viscosity\cite{read2009,read2011}
\begin{align}\label{lambda}
\eta_H=\frac{\lambda}{2}=\frac{\bar{s}\bar{\rho}}{2},
\end{align}
where $\bar{s}$ is the mean orbital spin per particle\cite{read2009,read2011} or guiding-center spin,\cite{HaldaneMetric,haldane2011,HaldaneFieldTheory} an intrinsic property of the FQH liquid related to the Wen-Zee shift $\c{S}$ on the sphere\cite{wen1992} by $\c{S}=2\bar{s}$. For the $\nu=1/q$ Laughlin state one has $\bar{s}=q/2$.\cite{read2009} 

This is a good place to note that the geometry of the local target space used in our field theoretic considerations is identical
to the geometry of the global parameter space used in the derivation of the Hall viscosity. Specifically the Hall viscosity arises
as a response to metric deformations that live in the coset space $SL(2,\mathbb{R})/SO(2)\cong SU(1,1)/U(1)$ and measures the adiabatic curvature of a
homogenous bundle over the same space.\cite{avron1995,levay1995,read2009,read2011} Thus it is not an accident that we reproduce the Hall viscosity at long wavelengths.

The spectrum of excitations in the isotropic phase contains the long-wavelength Kohn or cyclotron mode, which can be analyzed in either the direct or dual representations. In the dual representation (\ref{LdualFQHN}), long-wavelength modes correspond to the topologically trivial sector with no quasiparticle excitations $J^\mu_\textrm{vor}=J^\mu_\textrm{N}=0$. In this limit, the Maxwell-Chern-Simons theory for the statistical gauge field $a_\mu$ decouples from the nematic degrees of freedom, and is gapped\cite{deser1982} with a gap $\Delta=\kappa B$ corresponding to the cyclotron mode. In the direct representation (\ref{LagFluctQHN}), this mode is described by a Higgs-Chern-Simons theory which is the dual of the Maxwell-Chern-Simons theory. The Laughlin quasiparticle is a classical topological soliton of the direct Lagrangian (\ref{LagQHN}) corresponding to a $2\pi$ vortex in the phase field $\theta$, and its detailed solution can be obtained following the method of Ref.~\onlinecite{tafelmayer1993}.

The other collective mode at long wavelengths is the nematic mode, which is obtained by quantizing Eq.~(\ref{CollModesIsotropic}) for $\delta\bar{g}_{ab}=0$. The first term of $\c{L}_\textrm{N}$ gives the commutation relations $[\c{Q}(\b{r}),\c{Q}^*(\b{r}')]=-16\pi q\ell_B^2\bar{s}^{-1}\delta(\b{r}-\b{r}')$. The Hamiltonian $H_\textrm{N}=\int d^2r\,\c{H}_\textrm{N}$ can be diagonalized in momentum space $H_\textrm{N}=\int\frac{d^2k}{(2\pi)^2}E_\b{k}\beta_\b{k}^\dag\beta_\b{k}$ where $[\beta_\b{k},\beta_{\b{k}'}^\dag]=(2\pi)^2\delta(\b{k}-\b{k}')$ are canonically normalized boson operators and the mode spectrum is
\begin{align}\label{SpectrumIsotropic}
E_\b{k}=16\pi q\ell_B^2\bar{s}^{-1}\left(\frac{u_1}{2}(1-g)+\frac{\kappa_Q}{4}\b{k}^2\right),\hspace{5mm}g<1,
\end{align}
i.e., a gapped mode with quadratic dispersion. For $\kappa_Q<0$, the mode disperses downward at small $\b{k}$ and terms of higher order in $\b{k}$ are necessary to stabilize it. In this case, upon tuning the parameter $g$ the gap will collapse at a finite momentum $\b{k}\neq 0$, leading to charge density wave order (CDW). However, such a CDW instability will presumably be preempted by a first order transition.\cite{QHE} On the other hand, there are no fundamental reasons why microscopic interactions could not give the opposite sign $\kappa_Q>0$, in which case the gap $\Delta\propto(1-g)$ would collapse at $\b{k}=0$ as $g\rightarrow 1$. This is the regime where we expect our long-wavelength field theory to be valid, and we predict an instability to a nematic phase (assuming a continuous transition). Deep in the isotropic phase, the nematic mode (\ref{SpectrumIsotropic}) corresponds to fluctuations of 
the metric field $g_{ab}(\b{r},t)$ and should therefore be identified as the GMP mode. The equal-time structure factor $S(\b{k})=\langle[J^0(\b{k}),J^0(-\b{k})]\rangle$ is most easily calculated in the dual theory (\ref{LdualFQHN}). Varying $A_0$ sets the density $J^0$ equal to the flux of the dual gauge field $\b{a}$, and varying $a_0$ in the absence of vortices $J^0_\textrm{vor}=0$ sets this flux equal to $\nu$ times the nematic topological density $J^0_\textrm{N}$, hence $S(\b{k})=\nu^2\langle[J^0_\textrm{N}(\b{k}),J^0_\textrm{N}(-\b{k})]\rangle$. From Eq.~(\ref{alphaNcfinal}) and (\ref{JtopologicalNematic}), we find that only the second term in Eq.~(\ref{alphaNcfinal}) contributes at zero temperature and we have
\begin{align}
S(\b{k})\propto C^2(k\ell_B)^4,\hspace{5mm}
\b{k}\rightarrow 0,\nonumber
\end{align}
an important result of our work, which is in accordance with the GMP result.\cite{GMP} For trial wave functions restricted to the lowest Landau level, the coefficient of the $(k\ell_B)^4$ term in $S(\b{k})$ has a prescribed value,\cite{read2011} which would then fix our parameter $C$. In the presence of Landau level mixing, it is not clear to us that this coefficient should be universal and $C$ will depend on microscopic details.

\subsection{Nematic phase}

The nematic phase $g>1$ is a FQH state with spontaneously broken $SO(2)$ rotation symmetry and nonzero order parameter $\langle Q\rangle=\bar{Q}$. We first study the long-wavelength modes in the dual representation. Once again, in the absence of quasiparticles $J^\mu_\textrm{vor}=J^\mu_\textrm{N}=0$ the Maxwell-Chern-Simons theory for $a_\mu$ decouples from the nematic degrees of freedom and is gapped with gap $\Delta=\kappa B$ corresponding to the cyclotron mode. For the nematic sector, we write $Q=\bar{Q}+\chi$ with $\chi\ll\bar{Q}$ and expand $\c{L}_\textrm{N}$ in powers of $\chi$ and $\varphi$ to leading order (we can choose $\langle\varphi\rangle=0$ by $SO(2)$ symmetry). We obtain
\begin{eqnarray*}
\c{L}_\textrm{N}&=&\frac{\bar{s}\bar{\rho}}{2}\sqrt{g^2-1}\chi\partial_0\varphi-\frac{\kappa_Q}{4}\left[(\partial_i\chi)^2+(g^2-1)(\partial_i\varphi)^2\right]\nonumber\\
&&-\frac{u_1}{2}(g^2-1)\chi^2.
\end{eqnarray*}
The amplitude fluctuations $\chi$ can be integrated out,
\begin{eqnarray}\label{CollModesNematic}
\c{L}_\textrm{N}&=&\frac{(\bar{s}\bar{\rho})^2}{8}(g^2-1)\partial_0\varphi\frac{1}{u_1(g^2-1)-\frac{\kappa_Q}{2}\partial_i^2}\partial_0\varphi\nonumber\\
&&-\frac{\kappa_Q}{4}(g^2-1)(\partial_i\varphi)^2\nonumber\\
&=&\frac{(\bar{s}\bar{\rho})^2}{8u_1}(\partial_0\varphi)^2-\frac{\kappa_Q}{4}(g^2-1)(\partial_i\varphi)^2,
\end{eqnarray}
to leading order in a gradient expansion of $\varphi$.

Integrating out the density $\delta\rho$ and nematic $\chi$ amplitude fluctuations in the direct Lagrangian (\ref{LagFluctQHN}), we find that the nematic director fluctuations $\varphi$ decouple from the gauge fluctuations $\delta\alpha_\mu$ at long wavelengths and the Hall conductivity in the nematic phase remains quantized to $(1/q)(e^2/h)$. The long-wavelength spectrum contains the gapped cyclotron mode as in the isotropic phase (Sec.~\ref{sec:IsotropicPhase}), but also a gapless Goldstone mode with dispersion
\begin{align}\label{NematicGoldstone}
E_\b{k}=2\pi q\ell_B^2\bar{s}^{-1}\sqrt{2u_1\kappa_Q(g^2-1)}|\b{k}|,\hspace{5mm}g>1.
\end{align}
The velocity $v$ vanishes at the transition point $g=1$ as $v\sim(g-1)^{1/2}$, where the dispersion changes from linear to quadratic [Eq.~(\ref{SpectrumIsotropic})] in the isotropic phase. The Hall conductivity remains quantized in the nematic phase because the Goldstone mode is electrically neutral, just as in QH ferromagnetism.\cite{sondhi1993}

The nematic phase admits two types of topological defects: the Laughlin quasiparticle, which is the same as in Sec.~\ref{sec:IsotropicPhase}, and a nematic disclination which is a $2\pi$ vortex in the director angle $\varphi$. Indeed, although the nematic director (\ref{director}) changes sign under $\varphi\rightarrow\varphi+2\pi$, the nematic order parameter (\ref{Qab}) is invariant under $\hat{\b{d}}\rightarrow-\hat{\b{d}}$ and thus remains single-valued for a $2\pi$ vortex of $\varphi$. Mathematically, the target space for $\hat{\b{d}}$ is the real projective space $\mathbb{R}P^1=S^1/\mathbb{Z}_2$ which is topologically equivalent to $S^1$, i.e., $\pi_1(\mathbb{R}P^1)=\mathbb{Z}$ and nematic defects have the same homotopy classification as $U(1)$ vortices.\cite{DeGennes} However, the nematic defect exists only in the nematic phase where $\langle Q\rangle=\bar{Q}\neq 0$, whereas the Laughlin quasiparticle exists in both the isotropic and nematic phases. In contrast to the Laughlin quasiparticle, a single nematic disclination has an energy which diverges logarithmically with system size.\cite{DeGennes} It is accompanied by a flux of the gauge field $\delta\boldsymbol{\alpha}$ and thus carries electric charge. To the difference of the Laughlin quasiparticle, this electric charge is not quantized and depends on the value $\bar{Q}$ of the nematic order parameter. It is nonzero in the nematic phase and vanishes at the transition where $\bar{Q}$ vanishes. Given that the nematic disclination current $J^\mu_\textrm{N}$ is coupled to the statistical gauge field $a_\mu$ in the dual field theory (\ref{LdualFQHN}), the nematic disclinations have fractional statistics.

\subsection{Critical point: $z=2$}

Integrating out the non-dynamical fields $\delta\rho$ and $\delta\alpha_0$ in the direct representation (\ref{LagFluctQHN}), we obtain the critical theory as a sum of three terms,
\begin{align}
\c{L}=\c{L}(\c{Q})+\c{L}(\delta\alpha_\mu)
+\c{L}_\textrm{int}(\c{Q},\delta\alpha_\mu),\nonumber
\end{align}
which we expand about the Gaussian fixed point for $\c{Q}$, keeping all relevant and marginal terms. The nematic sector is described by
\begin{align}\label{LQz=2}
\mathcal{L}(\mathcal{Q})=-\frac{i\bar{s}\bar{\rho}}{4}\mathcal{Q}^*\partial_0\mathcal{Q}
-\frac{\tilde{\kappa}_Q}{4}|\partial_i\mathcal{Q}|^2
-\tilde{r}|\c{Q}|^2-\tilde{u}|\c{Q}|^4,
\end{align}
where $\tilde{\kappa}_Q=\kappa_Q+2\kappa\bar{\rho}C^2$, which implies a dynamic critical exponent $z=2$. The relevant mass term $\tilde{r}\propto(1-g)$ vanishes at the critical point, and the $|\c{Q}|^4$ term is marginal in 2D. Equation (\ref{LQz=2}) has been studied before as the Lagrangian of the dilute Bose gas;\cite{Sachdev} $d=2$ is the upper critical dimension and correlation functions acquire logarithmic corrections.\cite{prokofev2001} The gauge sector is described by
\begin{align}
\mathcal{L}(\delta\alpha_\mu)=\frac{1}{2\pi q}
\delta\dot{\alpha}_1\delta\alpha_2-\frac{\kappa\bar{\rho}}
{2}(\delta\alpha_i)^2,\nonumber
\end{align}
a Higgs-Chern-Simons theory which corresponds to a single massive scalar field $\phi\equiv\delta\alpha_1$ with conjugate momentum $\Pi=\partial\mathcal{L}/\partial\dot{\phi}=(1/2\pi q)\delta\alpha_2$ and Hamiltonian
\begin{align}
\mathcal{H}(\phi,\Pi)=\frac{(2\pi q)^2\kappa\bar{\rho}}{2}
\left(\Pi^2+\frac{1}{(2\pi q)^2}\phi^2\right).\nonumber
\end{align}
This theory is gapped with a gap $\Delta=2\pi q\kappa\bar{\rho}=\kappa B$ corresponding to the cyclotron mode. The coupling between nematic and gauge sectors is
\begin{align}
\c{L}_\textrm{int}(\c{Q},\delta\alpha_\mu)
=-\kappa\bar{\rho}C\epsilon_{ab}
\delta\alpha_c\partial_a\hat{Q}_{bc},\nonumber
\end{align}
which upon integrating out $\delta\alpha_c$ simply renormalizes the parameter $\tilde{\kappa}_Q$. Therefore the critical theory is Eq.~(\ref{LQz=2}), in the universality class of the 2D dilute Bose gas.

\section{Conclusions}
\label{sec:conclusion}

We have constructed a field theory of the isotropic-to-nematic transition in a Laughlin FQH liquid, assuming the transition is continuous. The basic ingredients of this field theory are the CS field $\alpha_\mu$ or its dual $a_\mu$ which describe the topological degrees of freedom, and the conventional nematic order parameter $\hat{Q}_{ab}$ which acquires a nonzero expectation value in the broken symmetry phase. To arrive at a prescription for coupling the nematic order parameter to the topological degrees of freedom, we seek inspiration in the problem of QH ferromagnetism. We propose two formulations which are the dual of each other in the field theory sense. In the direct formulation, we construct a gauge field $\alpha_\mu^\textrm{N}$ from the nematic order parameter $\hat{Q}_{ab}$. This new gauge field is then added to the CS gauge field $\alpha_\mu$, just as in the QH ferromagnetism problem where the extra gauge field is the $\mathbb{C}P^1$ gauge field built from the spin degrees of freedom. In the dual formulation, the nematic order parameter is used to construct a conserved topological current which is coupled to the dual gauge field. This current is the analog of the skyrmion current in QH ferromagnetism.

The first consequence of our field theory is that one can view an isotropic FQH liquid as a quantum disordered nematic. In this disordered phase, we obtain two gapped modes at long wavelengths: the Kohn mode and a mode corresponding to fluctuations of the nematic order parameter, this latter mode being identified with the GMP mode. Due to the coupling between the nematic order parameter and the CS and electromagnetic gauge fields, zero-point fluctuations of the nematic order parameter give a contribution $\propto k^4$ to the equal-time density structure factor $S(\b{k})$ as $\b{k}\rightarrow 0$. The isotropic phase admits one type of topological defect corresponding to the Laughlin quasiparticle with fractional charge and statistics, and the Hall conductivity is quantized as expected.

In the nematic phase, $SO(2)$ rotation symmetry is spontaneously broken by a nonzero order parameter $\langle\hat{Q}_{ab}\rangle\neq 0$. We find a linearly dispersing gapless Goldstone mode
with a velocity that vanishes upon approach to the quantum critical point.
Despite this gapless mode, the Hall conductivity remains quantized as in the isotropic phase because the gapless mode is electrically neutral. The Laughlin quasiparticle remains gapped through the phase transition and thus persists in the nematic phase. It is not involved in the transition in any essential way. However, the nematic phase supports additional topological defects: nematic disclinations with a logarithmically divergent energy. We expect these disclinations to proliferate above a Kosterlitz-Thouless temperature, where nematic order is destroyed.\cite{fradkin2000} The nematic disclinations carry electric charge, but unlike for Laughlin quasiparticles this charge is not quantized. At finite temperature below the Kosterlitz-Thouless temperature, we expect the loss of topological order but nematic order will persist.

Our theory predicts a critical point with $z=2$ scaling, which was also found in Ref.~\onlinecite{mulliganFQHN}. However, our two approaches differ widely. In Ref.~\onlinecite{mulliganFQHN}, the spontaneous breaking of rotation invariance occurs in the CS gauge field sector, whereas in our work it is described by the conventional nematic order parameter $\hat{Q}_{ab}$. Furthermore, in our approach the energy of the Kohn mode at zero momentum is set by the parameter $\kappa$ and is unaffected by the transition, allowing us to describe a Galilean invariant system.\cite{kohn1961} Our critical theory is in the universality class of the 2D dilute Bose gas. In 2D, this model is at its upper critical dimension and Gaussian $z=2$ scaling receives logarithmic corrections.

Interestingly, we find that our theory based on the nematic order parameter $\hat{Q}_{ab}(\b{r},t)$ and Haldane's recent field theory of the FQHE which involves a dynamical ``metric'' field $g_{ab}(\b{r},t)$ can be made to essentially agree if one identifies $g=\exp\hat{Q}$ in the sense of matrix exponentiation. [The reader may wish to revisit some fine print on this identification contained in our remarks following Eq.~(\ref{LagQHN}).] This leads us to identify the gapped nematic mode in the isotropic phase as the GMP mode of the Laughlin liquid. We predict that if the GMP mode can be made to collapse at zero momentum, either experimentally or in numerical studies of model Hamiltonians for the FQHE, an isotropic-to-nematic transition should occur (assuming the transition is continuous). For numerical studies, anisotropic versions of the Laughlin wave function\cite{haldane2011,yang2012,qiu2012,
yang2012b,wang2012,papic2013}  corresponding to $\langle g_{ab}\rangle\neq\delta_{ab}$ or, equivalently, $\langle\hat{Q}_{ab}\rangle\neq 0$, can be used as variational wave functions. However, to the difference of the numerical studies in Ref.~\onlinecite{qiu2012,yang2012b,wang2012,papic2013} where rotation symmetry was explicitly broken in the Hamiltonian, we suggest searching for model Hamiltonians which preserve rotation symmetry but with ground states which break rotation symmetry \emph{spontaneously} for some values of the parameters.

Finally, our prescription for coupling conventional order parameters to emergent gauge fields is a general framework for constructing effective field theories of topological phases (loosely defined) with conventional symmetry-breaking instabilities. Promising directions include the study of such instabilities in a $U(1)$ spin liquid or in non-Abelian FQH liquids, where the emergent gauge fields are non-Abelian and an even richer application of our ideas is possible.

\acknowledgments

We acknowledge B. A. Bernevig, E. Fradkin, S. Kachru, Z. Papi\'{c}, N. Read, N. Regnault, E. Witten, B. Yang, and especially F. D. M. Haldane for helpful discussions. This research was supported in part by the National Science Foundation under Grant No. DMR 10-06608 (SLS) and PHY-1005429 (BH and SLS), the U.S. Department of Energy under Grant No. AC02-76SF00515 (SAK), and the Simons Foundation (JM). YeJe Park was supported by Department of Energy, Office of Basic Energy Sciences through Grant No. \uppercase{DE-SC0002140}.

\bibliography{fqhn}

\begin{thebibliography}{67}
\expandafter\ifx\csname natexlab\endcsname\relax\def\natexlab#1{#1}\fi
\expandafter\ifx\csname bibnamefont\endcsname\relax
  \def\bibnamefont#1{#1}\fi
\expandafter\ifx\csname bibfnamefont\endcsname\relax
  \def\bibfnamefont#1{#1}\fi
\expandafter\ifx\csname citenamefont\endcsname\relax
  \def\citenamefont#1{#1}\fi
\expandafter\ifx\csname url\endcsname\relax
  \def\url#1{\texttt{#1}}\fi
\expandafter\ifx\csname urlprefix\endcsname\relax\def\urlprefix{URL }\fi
\providecommand{\bibinfo}[2]{#2}
\providecommand{\eprint}[2][]{\url{#2}}

\bibitem[{\citenamefont{Prange and Girvin}(1987)}]{QHE}
\bibinfo{author}{\bibfnamefont{R.~E.} \bibnamefont{Prange}} \bibnamefont{and}
  \bibinfo{author}{\bibfnamefont{S.~M.} \bibnamefont{Girvin}},
  \emph{\bibinfo{title}{The Quantum Hall Effect}}
  (\bibinfo{publisher}{Springer-Verlag}, \bibinfo{address}{New York},
  \bibinfo{year}{1987}).

\bibitem[{\citenamefont{Girvin and {MacDonald}}(1987)}]{girvin1987}
\bibinfo{author}{\bibfnamefont{S.~M.} \bibnamefont{Girvin}} \bibnamefont{and}
  \bibinfo{author}{\bibfnamefont{A.~H.} \bibnamefont{{MacDonald}}},
  \bibinfo{journal}{Phys. Rev. Lett.} \textbf{\bibinfo{volume}{58}},
  \bibinfo{pages}{1252} (\bibinfo{year}{1987}).

\bibitem[{zha()}]{zhang1989}
\bibinfo{note}{S. C. Zhang, T. H. Hansson, and S. A. Kivelson, Phys. Rev. Lett.
  {\bf 62}, 82 (1989); S. C. Zhang, Int. J. Mod. Phys. B {\bf 6}, 25 (1992).}

\bibitem[{\citenamefont{Lopez and Fradkin}(1991)}]{lopez1991}
\bibinfo{author}{\bibfnamefont{A.}~\bibnamefont{Lopez}} \bibnamefont{and}
  \bibinfo{author}{\bibfnamefont{E.}~\bibnamefont{Fradkin}},
  \bibinfo{journal}{Phys. Rev. B} \textbf{\bibinfo{volume}{44}},
  \bibinfo{pages}{5246} (\bibinfo{year}{1991}).

\bibitem[{\citenamefont{Haldane}(2011)}]{haldane2011}
\bibinfo{author}{\bibfnamefont{F.~D.~M.} \bibnamefont{Haldane}},
  \bibinfo{journal}{Phys. Rev. Lett.} \textbf{\bibinfo{volume}{107}},
  \bibinfo{pages}{116801} (\bibinfo{year}{2011}).

\bibitem[{Hal({\natexlab{a}})}]{HaldaneFieldTheory}
\bibinfo{note}{F. D. M. Haldane, seminar at the Rudolf Peierls Centre, Oxford
  University, Oxford, 23 June 2011.}

\bibitem[{unq()}]{unquantizedFQHN}
\bibinfo{note}{The simultaneous existence of topological order and nematicity
  distinguishes this phase from the unquantized nematic discussed previously in
  the QH regime, e.g., in Ref.~\onlinecite{lilly1999,wexler2001,
  FradkinKivelson,radzihovsky}.}

\bibitem[{mul()}]{mulliganFQHN}
\bibinfo{note}{M. Mulligan, C. Nayak, and S. Kachru, Phys. Rev. B {\bf 82},
  085102 (2010); {\bf 84}, 195124 (2011).}

\bibitem[{\citenamefont{Xia et~al.}(2011)\citenamefont{Xia, Eisenstein,
  Pfeiffer, and West}}]{xia2011}
\bibinfo{author}{\bibfnamefont{J.}~\bibnamefont{Xia}},
  \bibinfo{author}{\bibfnamefont{J.~P.} \bibnamefont{Eisenstein}},
  \bibinfo{author}{\bibfnamefont{L.~N.} \bibnamefont{Pfeiffer}},
  \bibnamefont{and} \bibinfo{author}{\bibfnamefont{K.~W.} \bibnamefont{West}},
  \bibinfo{journal}{Nature Phys.} \textbf{\bibinfo{volume}{7}},
  \bibinfo{pages}{845} (\bibinfo{year}{2011}).

\bibitem[{aba()}]{abanin2010}
\bibinfo{note}{D. A. Abanin, S. A. Parameswaran, S. A. Kivelson, and S. L.
  Sondhi, Phys. Rev. B {\bf 82}, 035428 (2010).}

\bibitem[{\citenamefont{Shkolnikov et~al.}(2005)\citenamefont{Shkolnikov,
  Misra, Bishop, De~Poortere, and Shayegan}}]{shkolnikov2005}
\bibinfo{author}{\bibfnamefont{Y.~P.} \bibnamefont{Shkolnikov}},
  \bibinfo{author}{\bibfnamefont{S.}~\bibnamefont{Misra}},
  \bibinfo{author}{\bibfnamefont{N.~C.} \bibnamefont{Bishop}},
  \bibinfo{author}{\bibfnamefont{E.~P.} \bibnamefont{De~Poortere}},
  \bibnamefont{and} \bibinfo{author}{\bibfnamefont{M.}~\bibnamefont{Shayegan}},
  \bibinfo{journal}{Phys. Rev. Lett.} \textbf{\bibinfo{volume}{95}},
  \bibinfo{pages}{066809} (\bibinfo{year}{2005}).

\bibitem[{\citenamefont{Gokmen and Shayegan}(2010)}]{gokmen2010}
\bibinfo{author}{\bibfnamefont{T.}~\bibnamefont{Gokmen}} \bibnamefont{and}
  \bibinfo{author}{\bibfnamefont{M.}~\bibnamefont{Shayegan}},
  \bibinfo{journal}{Phys. Rev. B} \textbf{\bibinfo{volume}{81}},
  \bibinfo{pages}{115336} (\bibinfo{year}{2010}).

\bibitem[{\citenamefont{{de Gennes} and Prost}(1993)}]{DeGennes}
\bibinfo{author}{\bibfnamefont{P.~G.} \bibnamefont{{de Gennes}}}
  \bibnamefont{and} \bibinfo{author}{\bibfnamefont{J.}~\bibnamefont{Prost}},
  \emph{\bibinfo{title}{The Physics of Liquid Crystals}}
  (\bibinfo{publisher}{Clarendon Press}, \bibinfo{address}{Oxford},
  \bibinfo{year}{1993}).

\bibitem[{\citenamefont{Sondhi et~al.}(1993)\citenamefont{Sondhi, Karlhede,
  Kivelson, and Rezayi}}]{sondhi1993}
\bibinfo{author}{\bibfnamefont{S.~L.} \bibnamefont{Sondhi}},
  \bibinfo{author}{\bibfnamefont{A.}~\bibnamefont{Karlhede}},
  \bibinfo{author}{\bibfnamefont{S.~A.} \bibnamefont{Kivelson}},
  \bibnamefont{and} \bibinfo{author}{\bibfnamefont{E.~H.}
  \bibnamefont{Rezayi}}, \bibinfo{journal}{Phys. Rev. B}
  \textbf{\bibinfo{volume}{47}}, \bibinfo{pages}{16419} (\bibinfo{year}{1993}).

\bibitem[{\citenamefont{Lee and Zhang}(1991)}]{lee1991}
\bibinfo{author}{\bibfnamefont{D.~H.} \bibnamefont{Lee}} \bibnamefont{and}
  \bibinfo{author}{\bibfnamefont{S.~C.} \bibnamefont{Zhang}},
  \bibinfo{journal}{Phys. Rev. Lett.} \textbf{\bibinfo{volume}{66}},
  \bibinfo{pages}{1220} (\bibinfo{year}{1991}).

\bibitem[{\citenamefont{Yang et~al.}(2012{\natexlab{a}})\citenamefont{Yang, Hu,
  Papi\'{c}, and Haldane}}]{yang2012}
\bibinfo{author}{\bibfnamefont{B.}~\bibnamefont{Yang}},
  \bibinfo{author}{\bibfnamefont{Z.-X.} \bibnamefont{Hu}},
  \bibinfo{author}{\bibfnamefont{Z.}~\bibnamefont{Papi\'{c}}},
  \bibnamefont{and} \bibinfo{author}{\bibfnamefont{F.~D.~M.}
  \bibnamefont{Haldane}}, \bibinfo{journal}{Phys. Rev. Lett.}
  \textbf{\bibinfo{volume}{108}}, \bibinfo{pages}{256807}
  (\bibinfo{year}{2012}{\natexlab{a}}).

\bibitem[{GMP()}]{GMP}
\bibinfo{note}{S. M. Girvin, A. H. MacDonald, and P. M. Platzman, Phys. Rev.
  Lett. {\bf 54}, 581 (1985); Phys. Rev. B {\bf 33}, 2481 (1986).}

\bibitem[{\citenamefont{Wen}(2004)}]{WenBook}
\bibinfo{author}{\bibfnamefont{X.~G.} \bibnamefont{Wen}},
  \emph{\bibinfo{title}{Quantum Field Theory of Many-Body Systems: From the
  Origin of Sound to an Origin of Light and Electrons}}
  (\bibinfo{publisher}{Oxford University Press}, \bibinfo{address}{Oxford},
  \bibinfo{year}{2004}).

\bibitem[{Hal({\natexlab{b}})}]{HaldaneGap}
\bibinfo{note}{F. D. M. Haldane, Phys. Lett. A {\bf 93}, 464 (1983); Phys. Rev.
  Lett. {\bf 50}, 1153 (1983).}

\bibitem[{\citenamefont{Gu and Wen}(2009)}]{gu2009}
\bibinfo{author}{\bibfnamefont{Z.~C.} \bibnamefont{Gu}} \bibnamefont{and}
  \bibinfo{author}{\bibfnamefont{X.~G.} \bibnamefont{Wen}},
  \bibinfo{journal}{Phys. Rev. B} \textbf{\bibinfo{volume}{80}},
  \bibinfo{pages}{155131} (\bibinfo{year}{2009}).

\bibitem[{\citenamefont{Pollmann et~al.}(2012)\citenamefont{Pollmann, Berg,
  Turner, and Oshikawa}}]{pollmann2012}
\bibinfo{author}{\bibfnamefont{F.}~\bibnamefont{Pollmann}},
  \bibinfo{author}{\bibfnamefont{E.}~\bibnamefont{Berg}},
  \bibinfo{author}{\bibfnamefont{A.~M.} \bibnamefont{Turner}},
  \bibnamefont{and} \bibinfo{author}{\bibfnamefont{M.}~\bibnamefont{Oshikawa}},
  \bibinfo{journal}{Phys. Rev. B} \textbf{\bibinfo{volume}{85}},
  \bibinfo{pages}{075125} (\bibinfo{year}{2012}).

\bibitem[{\citenamefont{Affleck}(1986)}]{affleck1986}
\bibinfo{author}{\bibfnamefont{I.}~\bibnamefont{Affleck}},
  \bibinfo{journal}{Nucl. Phys. B} \textbf{\bibinfo{volume}{265}},
  \bibinfo{pages}{409} (\bibinfo{year}{1986}).

\bibitem[{\citenamefont{Musaelian and Joynt}(1996)}]{musaelian1996}
\bibinfo{author}{\bibfnamefont{K.}~\bibnamefont{Musaelian}} \bibnamefont{and}
  \bibinfo{author}{\bibfnamefont{R.}~\bibnamefont{Joynt}}, \bibinfo{journal}{J.
  Phys. Condens. Matter} \textbf{\bibinfo{volume}{8}}, \bibinfo{pages}{L105}
  (\bibinfo{year}{1996}).

\bibitem[{\citenamefont{Balents}(1996)}]{balentshexatic}
\bibinfo{author}{\bibfnamefont{L.}~\bibnamefont{Balents}},
  \bibinfo{journal}{Europhys. Lett.} \textbf{\bibinfo{volume}{33}},
  \bibinfo{pages}{291} (\bibinfo{year}{1996}).

\bibitem[{\citenamefont{Fisher and Lee}(1989)}]{fisher1989}
\bibinfo{author}{\bibfnamefont{M.~P.~A.} \bibnamefont{Fisher}}
  \bibnamefont{and} \bibinfo{author}{\bibfnamefont{D.~H.} \bibnamefont{Lee}},
  \bibinfo{journal}{Phys. Rev. B} \textbf{\bibinfo{volume}{39}},
  \bibinfo{pages}{2756} (\bibinfo{year}{1989}).

\bibitem[{\citenamefont{Tafelmayer}(1993)}]{tafelmayer1993}
\bibinfo{author}{\bibfnamefont{R.}~\bibnamefont{Tafelmayer}},
  \bibinfo{journal}{Nucl. Phys. B} \textbf{\bibinfo{volume}{396}},
  \bibinfo{pages}{386} (\bibinfo{year}{1993}).

\bibitem[{\citenamefont{Kohn}(1961)}]{kohn1961}
\bibinfo{author}{\bibfnamefont{W.}~\bibnamefont{Kohn}}, \bibinfo{journal}{Phys.
  Rev.} \textbf{\bibinfo{volume}{123}}, \bibinfo{pages}{1242}
  (\bibinfo{year}{1961}).

\bibitem[{\citenamefont{Sachdev}(2011)}]{Sachdev}
\bibinfo{author}{\bibfnamefont{S.}~\bibnamefont{Sachdev}},
  \emph{\bibinfo{title}{Quantum Phase Transitions}}
  (\bibinfo{publisher}{Cambridge University Press},
  \bibinfo{address}{Cambridge}, \bibinfo{year}{2011}).

\bibitem[{nak()}]{nakahara}
\bibinfo{note}{M. Nakahara, \emph{Geometry, Topology and Physics} (Institute of
  Physics Publishing, London, 1990).}

\bibitem[{\citenamefont{Lee and Kane}(1990)}]{lee1990}
\bibinfo{author}{\bibfnamefont{D.~H.} \bibnamefont{Lee}} \bibnamefont{and}
  \bibinfo{author}{\bibfnamefont{C.~L.} \bibnamefont{Kane}},
  \bibinfo{journal}{Phys. Rev. Lett.} \textbf{\bibinfo{volume}{64}},
  \bibinfo{pages}{1313} (\bibinfo{year}{1990}).

\bibitem[{\citenamefont{Wen}(1989)}]{wen1989}
\bibinfo{author}{\bibfnamefont{X.~G.} \bibnamefont{Wen}},
  \bibinfo{journal}{Phys. Rev. B} \textbf{\bibinfo{volume}{40}},
  \bibinfo{pages}{7387} (\bibinfo{year}{1989}).

\bibitem[{\citenamefont{Wen and Niu}(1990)}]{wen1990}
\bibinfo{author}{\bibfnamefont{X.~G.} \bibnamefont{Wen}} \bibnamefont{and}
  \bibinfo{author}{\bibfnamefont{Q.}~\bibnamefont{Niu}},
  \bibinfo{journal}{Phys. Rev. B} \textbf{\bibinfo{volume}{41}},
  \bibinfo{pages}{9377} (\bibinfo{year}{1990}).

\bibitem[{\citenamefont{Barrett et~al.}(1995)\citenamefont{Barrett, Dabbagh,
  Pfeiffer, West, and Tycko}}]{barrett1995}
\bibinfo{author}{\bibfnamefont{S.~E.} \bibnamefont{Barrett}},
  \bibinfo{author}{\bibfnamefont{G.}~\bibnamefont{Dabbagh}},
  \bibinfo{author}{\bibfnamefont{L.~N.} \bibnamefont{Pfeiffer}},
  \bibinfo{author}{\bibfnamefont{K.~W.} \bibnamefont{West}}, \bibnamefont{and}
  \bibinfo{author}{\bibfnamefont{R.}~\bibnamefont{Tycko}},
  \bibinfo{journal}{Phys. Rev. Lett.} \textbf{\bibinfo{volume}{74}},
  \bibinfo{pages}{5112} (\bibinfo{year}{1995}).

\bibitem[{\citenamefont{Fradkin}(1991)}]{Fradkin}
\bibinfo{author}{\bibfnamefont{E.}~\bibnamefont{Fradkin}},
  \emph{\bibinfo{title}{Field Theories of Condensed Matter Systems}}
  (\bibinfo{publisher}{Addison-Wesley}, \bibinfo{address}{Redwood City},
  \bibinfo{year}{1991}).

\bibitem[{\citenamefont{Sakurai}(1994)}]{Sakurai}
\bibinfo{author}{\bibfnamefont{J.~J.} \bibnamefont{Sakurai}},
  \emph{\bibinfo{title}{Modern Quantum Mechanics}}
  (\bibinfo{publisher}{Addison-Wesley}, \bibinfo{address}{Reading},
  \bibinfo{year}{1994}).

\bibitem[{\citenamefont{Ryder}(1980)}]{ryder1980}
\bibinfo{author}{\bibfnamefont{L.~H.} \bibnamefont{Ryder}},
  \bibinfo{journal}{J. Phys. A} \textbf{\bibinfo{volume}{13}},
  \bibinfo{pages}{437} (\bibinfo{year}{1980}).

\bibitem[{\citenamefont{Wilczek and Zee}(1983)}]{wilczek1983}
\bibinfo{author}{\bibfnamefont{F.}~\bibnamefont{Wilczek}} \bibnamefont{and}
  \bibinfo{author}{\bibfnamefont{A.}~\bibnamefont{Zee}},
  \bibinfo{journal}{Phys. Rev. Lett.} \textbf{\bibinfo{volume}{51}},
  \bibinfo{pages}{2250} (\bibinfo{year}{1983}).

\bibitem[{bel()}]{belavin1975}
\bibinfo{note}{A. A. Belavin and A. M. Polyakov, Pis'ma Zh. Eksp. Teor. Fiz.
  {\bf 22}, 503 (1975) [JETP Lett. {\bf 22}, 245 (1975)].}

\bibitem[{\citenamefont{Eichenherr}(1978)}]{eichenherr1978}
\bibinfo{author}{\bibfnamefont{H.}~\bibnamefont{Eichenherr}},
  \bibinfo{journal}{Nucl. Phys. B} \textbf{\bibinfo{volume}{146}},
  \bibinfo{pages}{215} (\bibinfo{year}{1978}).

\bibitem[{\citenamefont{{D'Adda} et~al.}(1978)\citenamefont{{D'Adda},
  {L\"{u}scher}, and Di~Vecchia}}]{dadda1978}
\bibinfo{author}{\bibfnamefont{A.}~\bibnamefont{{D'Adda}}},
  \bibinfo{author}{\bibfnamefont{M.}~\bibnamefont{{L\"{u}scher}}},
  \bibnamefont{and}
  \bibinfo{author}{\bibfnamefont{P.}~\bibnamefont{Di~Vecchia}},
  \bibinfo{journal}{Nucl. Phys. B} \textbf{\bibinfo{volume}{146}},
  \bibinfo{pages}{63} (\bibinfo{year}{1978}).

\bibitem[{\citenamefont{Witten}(1979)}]{witten1979}
\bibinfo{author}{\bibfnamefont{E.}~\bibnamefont{Witten}},
  \bibinfo{journal}{Nucl. Phys. B} \textbf{\bibinfo{volume}{149}},
  \bibinfo{pages}{285} (\bibinfo{year}{1979}).

\bibitem[{\citenamefont{Auerbach}(1994)}]{Auerbach}
\bibinfo{author}{\bibfnamefont{A.}~\bibnamefont{Auerbach}},
  \emph{\bibinfo{title}{Interacting Electrons and Quantum Magnetism}}
  (\bibinfo{publisher}{Springer}, \bibinfo{address}{New York},
  \bibinfo{year}{1994}).

\bibitem[{\citenamefont{Balazs and Voros}(1986)}]{balazs1986}
\bibinfo{author}{\bibfnamefont{N.}~\bibnamefont{Balazs}} \bibnamefont{and}
  \bibinfo{author}{\bibfnamefont{A.}~\bibnamefont{Voros}},
  \bibinfo{journal}{Phys. Rep.} \textbf{\bibinfo{volume}{143}},
  \bibinfo{pages}{109} (\bibinfo{year}{1986}).

\bibitem[{Not({\natexlab{a}})}]{NoteQHIN}
\bibinfo{note}{For a study of Ising nematic order in a QH system, see
  Ref.~\onlinecite{abanin2010}.}

\bibitem[{Not({\natexlab{b}})}]{NoteLorentzSO21}
\bibinfo{note}{In the QH ferromagnetism problem, we were only interested in
  smooth rotations of $\b{n}$, hence we only kept the connected $SO(3)$
  component of the full $O(3)$ symmetry group, which includes inversions.
  Likewise, in the QH nematics problem we are only interested in smooth
  deformations of $Q_{ab}$ and $g_{ab}$, hence we only keep the connected
  $SO^+(2,1)$ ``orthochronous'' component of the full $SO(2,1)$ Lorentz group.
  In both cases, the Lie algebras of the full group and its connected component
  are isomorphic.}

\bibitem[{\citenamefont{Gilmore}(2008)}]{Gilmore}
\bibinfo{author}{\bibfnamefont{R.}~\bibnamefont{Gilmore}},
  \emph{\bibinfo{title}{Lie Groups, Physics, and Geometry}}
  (\bibinfo{publisher}{Cambridge University Press},
  \bibinfo{address}{Cambridge}, \bibinfo{year}{2008}).

\bibitem[{\citenamefont{Witten}(1988)}]{witten1988}
\bibinfo{author}{\bibfnamefont{E.}~\bibnamefont{Witten}},
  \bibinfo{journal}{Nucl. Phys. B} \textbf{\bibinfo{volume}{311}},
  \bibinfo{pages}{46} (\bibinfo{year}{1988}).

\bibitem[{Hal({\natexlab{c}})}]{HaldaneMetric}
\bibinfo{note}{F. D. M. Haldane, e-print arXiv:0906.1854 (unpublished);
  1112.0990 (unpublished).}

\bibitem[{\citenamefont{Jackson}(1962)}]{jackson}
\bibinfo{author}{\bibfnamefont{J.~D.} \bibnamefont{Jackson}},
  \emph{\bibinfo{title}{Classical Electrodynamics}}
  (\bibinfo{publisher}{Wiley}, \bibinfo{address}{New York},
  \bibinfo{year}{1962}).

\bibitem[{\citenamefont{Read}(2009)}]{read2009}
\bibinfo{author}{\bibfnamefont{N.}~\bibnamefont{Read}}, \bibinfo{journal}{Phys.
  Rev. B} \textbf{\bibinfo{volume}{79}}, \bibinfo{pages}{045308}
  (\bibinfo{year}{2009}).

\bibitem[{\citenamefont{Read and Rezayi}(2011)}]{read2011}
\bibinfo{author}{\bibfnamefont{N.}~\bibnamefont{Read}} \bibnamefont{and}
  \bibinfo{author}{\bibfnamefont{E.~H.} \bibnamefont{Rezayi}},
  \bibinfo{journal}{Phys. Rev. B} \textbf{\bibinfo{volume}{84}},
  \bibinfo{pages}{085316} (\bibinfo{year}{2011}).

\bibitem[{\citenamefont{Avron et~al.}(1995)\citenamefont{Avron, Seiler, and
  Zograf}}]{avron1995}
\bibinfo{author}{\bibfnamefont{J.~E.} \bibnamefont{Avron}},
  \bibinfo{author}{\bibfnamefont{R.}~\bibnamefont{Seiler}}, \bibnamefont{and}
  \bibinfo{author}{\bibfnamefont{P.~G.} \bibnamefont{Zograf}},
  \bibinfo{journal}{Phys. Rev. Lett.} \textbf{\bibinfo{volume}{75}},
  \bibinfo{pages}{697} (\bibinfo{year}{1995}).

\bibitem[{\citenamefont{Bradlyn et~al.}(2012)\citenamefont{Bradlyn, Goldstein,
  and Read}}]{bradlyn2012}
\bibinfo{author}{\bibfnamefont{B.}~\bibnamefont{Bradlyn}},
  \bibinfo{author}{\bibfnamefont{M.}~\bibnamefont{Goldstein}},
  \bibnamefont{and} \bibinfo{author}{\bibfnamefont{N.}~\bibnamefont{Read}},
  \bibinfo{journal}{Phys. Rev. B} \textbf{\bibinfo{volume}{86}},
  \bibinfo{pages}{245309} (\bibinfo{year}{2012}).

\bibitem[{\citenamefont{Hoyos and Son}(2012)}]{hoyos2012}
\bibinfo{author}{\bibfnamefont{C.}~\bibnamefont{Hoyos}} \bibnamefont{and}
  \bibinfo{author}{\bibfnamefont{D.~T.} \bibnamefont{Son}},
  \bibinfo{journal}{Phys. Rev. Lett.} \textbf{\bibinfo{volume}{108}},
  \bibinfo{pages}{066805} (\bibinfo{year}{2012}).

\bibitem[{\citenamefont{Wen and Zee}(1992)}]{wen1992}
\bibinfo{author}{\bibfnamefont{X.~G.} \bibnamefont{Wen}} \bibnamefont{and}
  \bibinfo{author}{\bibfnamefont{A.}~\bibnamefont{Zee}},
  \bibinfo{journal}{Phys. Rev. Lett.} \textbf{\bibinfo{volume}{69}},
  \bibinfo{pages}{953} (\bibinfo{year}{1992}).

\bibitem[{\citenamefont{{L\'{e}vay}}(1995)}]{levay1995}
\bibinfo{author}{\bibfnamefont{P.}~\bibnamefont{{L\'{e}vay}}},
  \bibinfo{journal}{J. Math. Phys.} \textbf{\bibinfo{volume}{36}},
  \bibinfo{pages}{2792} (\bibinfo{year}{1995}).

\bibitem[{\citenamefont{Deser et~al.}(1982)\citenamefont{Deser, Jackiw, and
  Templeton}}]{deser1982}
\bibinfo{author}{\bibfnamefont{S.}~\bibnamefont{Deser}},
  \bibinfo{author}{\bibfnamefont{R.}~\bibnamefont{Jackiw}}, \bibnamefont{and}
  \bibinfo{author}{\bibfnamefont{S.}~\bibnamefont{Templeton}},
  \bibinfo{journal}{Phys. Rev. Lett.} \textbf{\bibinfo{volume}{48}},
  \bibinfo{pages}{975} (\bibinfo{year}{1982}).

\bibitem[{\citenamefont{Prokof'ev et~al.}(2001)\citenamefont{Prokof'ev,
  Ruebenacker, and Svistunov}}]{prokofev2001}
\bibinfo{author}{\bibfnamefont{N.}~\bibnamefont{Prokof'ev}},
  \bibinfo{author}{\bibfnamefont{O.}~\bibnamefont{Ruebenacker}},
  \bibnamefont{and}
  \bibinfo{author}{\bibfnamefont{B.}~\bibnamefont{Svistunov}},
  \bibinfo{journal}{Phys. Rev. Lett.} \textbf{\bibinfo{volume}{87}},
  \bibinfo{pages}{270402} (\bibinfo{year}{2001}).

\bibitem[{\citenamefont{Fradkin et~al.}(2000)\citenamefont{Fradkin, Kivelson,
  Manousakis, and Nho}}]{fradkin2000}
\bibinfo{author}{\bibfnamefont{E.}~\bibnamefont{Fradkin}},
  \bibinfo{author}{\bibfnamefont{S.~A.} \bibnamefont{Kivelson}},
  \bibinfo{author}{\bibfnamefont{E.}~\bibnamefont{Manousakis}},
  \bibnamefont{and} \bibinfo{author}{\bibfnamefont{K.}~\bibnamefont{Nho}},
  \bibinfo{journal}{Phys. Rev. Lett.} \textbf{\bibinfo{volume}{84}},
  \bibinfo{pages}{1982} (\bibinfo{year}{2000}).

\bibitem[{\citenamefont{Qiu et~al.}(2012)\citenamefont{Qiu, Haldane, Wan, Yang,
  and Yi}}]{qiu2012}
\bibinfo{author}{\bibfnamefont{R.-Z.} \bibnamefont{Qiu}},
  \bibinfo{author}{\bibfnamefont{F.~D.~M.} \bibnamefont{Haldane}},
  \bibinfo{author}{\bibfnamefont{X.}~\bibnamefont{Wan}},
  \bibinfo{author}{\bibfnamefont{K.}~\bibnamefont{Yang}}, \bibnamefont{and}
  \bibinfo{author}{\bibfnamefont{S.}~\bibnamefont{Yi}}, \bibinfo{journal}{Phys.
  Rev. B} \textbf{\bibinfo{volume}{85}}, \bibinfo{pages}{115308}
  (\bibinfo{year}{2012}).

\bibitem[{\citenamefont{Yang et~al.}(2012{\natexlab{b}})\citenamefont{Yang,
  Papi\'{c}, Rezayi, Bhatt, and Haldane}}]{yang2012b}
\bibinfo{author}{\bibfnamefont{B.}~\bibnamefont{Yang}},
  \bibinfo{author}{\bibfnamefont{Z.}~\bibnamefont{Papi\'{c}}},
  \bibinfo{author}{\bibfnamefont{E.~H.} \bibnamefont{Rezayi}},
  \bibinfo{author}{\bibfnamefont{R.~N.} \bibnamefont{Bhatt}}, \bibnamefont{and}
  \bibinfo{author}{\bibfnamefont{F.~D.~M.} \bibnamefont{Haldane}},
  \bibinfo{journal}{Phys. Rev. B} \textbf{\bibinfo{volume}{85}},
  \bibinfo{pages}{165318} (\bibinfo{year}{2012}{\natexlab{b}}).

\bibitem[{\citenamefont{Wang et~al.}(2012)\citenamefont{Wang, Narayanan, Wan,
  and Zhang}}]{wang2012}
\bibinfo{author}{\bibfnamefont{H.}~\bibnamefont{Wang}},
  \bibinfo{author}{\bibfnamefont{R.}~\bibnamefont{Narayanan}},
  \bibinfo{author}{\bibfnamefont{X.}~\bibnamefont{Wan}}, \bibnamefont{and}
  \bibinfo{author}{\bibfnamefont{F.~C.} \bibnamefont{Zhang}},
  \bibinfo{journal}{Phys. Rev. B} \textbf{\bibinfo{volume}{86}},
  \bibinfo{pages}{035122} (\bibinfo{year}{2012}).

\bibitem[{\citenamefont{Papi\'{c}}(2013)}]{papic2013}
\bibinfo{author}{\bibfnamefont{Z.}~\bibnamefont{Papi\'{c}}},
  \bibinfo{journal}{Phys. Rev. B} \textbf{\bibinfo{volume}{87}},
  \bibinfo{pages}{245315} (\bibinfo{year}{2013}).

\bibitem[{lil()}]{lilly1999}
\bibinfo{note}{M. P. Lilly, K. B. Cooper, J. P. Eisenstein, L. N. Pfeiffer, and
  K. W. West, Phys. Rev. Lett. {\bf 82}, 394 (1999); {\bf 83}, 824 (1999).}

\bibitem[{\citenamefont{Wexler and Dorsey}(2001)}]{wexler2001}
\bibinfo{author}{\bibfnamefont{C.}~\bibnamefont{Wexler}} \bibnamefont{and}
  \bibinfo{author}{\bibfnamefont{A.~T.} \bibnamefont{Dorsey}},
  \bibinfo{journal}{Phys. Rev. B} \textbf{\bibinfo{volume}{64}},
  \bibinfo{pages}{115312} (\bibinfo{year}{2001}).

\bibitem[{\citenamefont{Fradkin and Kivelson}(1999)}]{FradkinKivelson}
\bibinfo{author}{\bibfnamefont{E.}~\bibnamefont{Fradkin}} \bibnamefont{and}
  \bibinfo{author}{\bibfnamefont{S.~A.} \bibnamefont{Kivelson}},
  \bibinfo{journal}{Phys. Rev. B} \textbf{\bibinfo{volume}{59}},
  \bibinfo{pages}{8065} (\bibinfo{year}{1999}).

\bibitem[{\citenamefont{Radzihovsky and Dorsey}(2002)}]{radzihovsky}
\bibinfo{author}{\bibfnamefont{L.}~\bibnamefont{Radzihovsky}} \bibnamefont{and}
  \bibinfo{author}{\bibfnamefont{A.~T.} \bibnamefont{Dorsey}},
  \bibinfo{journal}{Phys. Rev. Lett.} \textbf{\bibinfo{volume}{88}},
  \bibinfo{pages}{216802} (\bibinfo{year}{2002}).

\end{thebibliography}
\end{document}